%% file: CPAFM.tex
\documentclass[12pt]{article}

\renewcommand{\baselinestretch}{1.66}
\usepackage{amsmath}
\usepackage{amssymb}

\usepackage{array}

\usepackage{epsfig}
\usepackage{endfloat}
\nomarkersintext 

\usepackage{vmargin}
\setpapersize{USletter}
\setmarginsrb{1in}{0.5in}{1in}{0.5in}{0mm}{11mm}{0mm}{11mm}

\makeatletter
\renewcommand{\section}{\@startsection
	{section}
	{1}
	{\z@}
	{-3.5ex \@plus -1ex \@minus -.2ex}
	{2.3ex \@plus.2ex}
	{\normalfont\large\bfseries}
}
\makeatother
\makeatletter
\renewenvironment{abstract}{%
	\begin{center}%
    {\normalfont\large\bfseries \abstractname\vspace{-.5em}\vspace{\z@}}%
        \end{center}}{}
\makeatother

\newcommand{\foreign}{\em}
\newcommand{\sepCaption}[1]{\caption[#1]{}}
\newcommand{\etal}{{\foreign{et al.}}}
\newcommand{\eg}{{e.g.}}
\newcommand{\ie}{{i.e.}}

\renewcommand{\d}[1]{\,\mathrm{d}#1\,}
\newcommand{\re}{\mathop{\mathrm{Re}}}

\newcommand{\e}{\mathop{\mathrm{e}}\nolimits}
\newcommand{\iunit}{\mathrm{i}}

\newcommand{\conjugate}[1]{{#1}^*}

\newcommand{\Naturals}{\mathbb{N}}

\newlength{\figboxwidth} 

\newcommand{\nought}{\circ}
\newcommand{\vecz}{\mathbf z}
\newcommand{\vecx}{\mathbf x}
\newcommand{\vecq}{\mathbf q}
\newcommand{\vecr}{\mathbf r}
\newcommand{\dvecx}{\d{\vecx}}
\newcommand{\del}{\nabla}
\newcommand{\htilde}{\tilde{h}}
\newcommand{\vecm}{\mathbf m}
\newcommand{\vecM}{\mathbf M}
\newcommand{\stagMagMoment}{\vecm^{s}}
\newcommand{\magMoment}{\vecm}
\newcommand{\stagMag}{\vecM^{s}}
\newcommand{\magnetization}{\vecM}
\newcommand{\opO}{\mathrm{O}}
\newcommand{\parity}{\epsilon}
\newcommand{\aperp}{a_{\perp}}
\newcommand{\dperp}{d^{\perp}}
\newcommand{\Escript}{\mathrm E}
\newcommand{\Oscript}{\mathrm O}
\newcommand{\evenv}[1]{\left(#1\right)_{\Escript}}
\newcommand{\oddv}[1]{\left(#1\right)_{\Oscript}}
\newcommand{\DzE}{\Delta\vecz^{\Escript}}
\newcommand{\DzO}{\Delta\vecz^{\Oscript}}
\newcommand{\vech}{\mathbf h}
\newcommand{\tensorK}{\mathbf K}

\newcommand{\KE}{K^{\Escript}}
\newcommand{\KO}{K^{\Oscript}}
\newcommand{\tensorC}{\mathbf C}
\newcommand{\vechE}{\vech^{\Escript}}
\newcommand{\vechO}{\vech^{\Oscript}}
\newcommand{\veca}{\mathbf a}
\newcommand{\vecaE}{\veca^{\Escript}}
\newcommand{\vecaO}{\veca^{\Oscript}}
\newcommand{\vecaperp}{\veca^{\perp}}
\newcommand{\vecG}{\mathbf G}
\newcommand{\GE}{\vecG^{\Escript}}
\newcommand{\GO}{\vecG^{\Oscript}}
\newcommand{\mixscript}{\mathrm{mix}}
\newcommand{\Gmix}{\vecG^{\mixscript}}
\newcommand{\minscript}{\mathrm{min}}
\newcommand{\Gmin}{\vecG_{\minscript}}
\newcommand{\etaE}{\eta^{\Escript}}
\newcommand{\etaO}{\eta^{\Oscript}}
\newcommand{\etamix}{\eta^{\mixscript}}
\newcommand{\etamin}{\eta_{\minscript}}
\newcommand{\etarat}{\etaE/\etaO}
\newcommand{\etaM}{\eta_{\mathrm M}}
\newcommand{\etaS}{\eta_{\mathrm S}}
\newcommand{\lockscript}{\mathrm{lock}}
\newcommand{\flock}{f_{\lockscript}}
\newcommand{\Glock}{\vecG_{\lockscript}}
\newcommand{\etalock}{\eta_{\lockscript}}
\newcommand{\GlockE}{\Glock^{\Escript}}
\newcommand{\GlockO}{\Glock^{\Oscript}}
\newcommand{\Glockmix}{\Glock^{\mixscript}}
\newcommand{\etalockE}{\etalock^{\Escript}}
\newcommand{\etalockO}{\etalock^{\Oscript}}
\newcommand{\etalockmix}{\etalock^{\mixscript}}
\newcommand{\const}{\mathrm{const}}

\newcommand{\vecheven}{\vech_{\mathrm{even}}}
\newcommand{\vechodd}{\vech_{\mathrm{odd}}}
\newcommand{\yeff}{y_{\mathrm{eff}}}
\newcommand{\vecb}{\mathbf b}
\newcommand{\qcrit}{q_{\mathrm c}}
\newcommand{\paro}{\parity_{0}}
\newcommand{\sig}[1]{\sigma^{(#1)}}

\newenvironment{mysubequations}{\begin{subequations}\openup 2.5\jot}%
		{\end{subequations}}

\usepackage{kenjmps} 

\makeatletter

\renewcommand{\@pnumwidth}{0 pt}
\renewcommand{\@dotsep}{5000}
\renewcommand{\l@figure}[2]{\@dottedtocline{1}{1.5em}{2.3em}{Fig.~#1}{}}
\renewcommand{\l@table}[2]{\@dottedtocline{1}{1.5em}{2.3em}{Table~#1}{}}
\makeatother

\numberwithin{equation}{section}

\begin{document}
\nocite{KondevHenley:1995a} 
\pagenumbering{arabic}  
\include{CPAFMabstract}
\include{CPAFM1}
\include{CPAFM2}

\include{CPAFM3}
\include{CPAFM4}

\include{CPAFM5}
\include{CPAFM6}

\appendix\newpage
\makeatletter
\newcommand{\sectionappendixword}{APPENDIX }
\newcommand{\subsectionappendixword}{}
\newcommand{\sectionappendixcolon}{:}
\newcommand{\subsectionappendixcolon}{}
\renewcommand{\@seccntformat}[1]
	{{\csname #1appendixword\endcsname}{\csname the#1\endcsname}%
	{\csname #1appendixcolon\endcsname}\hspace{0.5em}}
\makeatother
\numberwithin{equation}{section}
\bibliographystyle{kenjmps}
\bibliography{abbreviations,heightrefs}
\include{CPAFMappA}

\include{CPAFMappB}
\include{CPAFMappC}
\end{document}

%% file: CPAFMabstract.tex
\renewcommand{\baselinestretch}{1.}
\title{A constrained Potts antiferromagnet model 
with an interface representation}

\author{J. K. BURTON JR. 
and C. L. HENLEY \\
Department of Physics, Cornell University \\
Ithaca, NY 14853-2501, U. S. A.}

\date{}
\maketitle

\noindent{}

\renewcommand{\baselinestretch}{1.66} 

\begin{abstract}
We define a four-state Potts model ensemble on the square 
lattice, with the constraints that neighboring spins must have 
different values, and that no plaquette may contain all four states.  
The spin configurations may be mapped into those of a 2-dimensional 
interface in a 2+5 dimensional space.  If this interface is in a 
Gaussian rough phase (as is the case for most other models with such a 
mapping), then the spin correlations are critical and their exponents 
can be related to the stiffness governing the interface fluctuations.  
Results of our Monte Carlo simulations show height fluctuations with 
an anomalous dependence on wavevector, intermediate between the 
behaviors expected in a rough phase and in a smooth phase; we argue 
that the smooth phase (which would imply long-range spin order) is the 
best interpretation.
\end{abstract}

%% file: CPAFM1.tex
\section{Introduction} \label{sec:1}

A number of discrete spin models with highly degenerate ground 
states have critical ground state ensembles: these include the 
triangular Ising antiferromagnet with spin 1/2
\cite{Wannier:1950,NienhuisEtal:1984} or with general spin 
\cite{LipowskiEtal:1995,ZengHenley:1997}, the three-state antiferromagnetic 
Potts model on the square lattice or equivalently the six-vertex 
model 
\cite{LiebWu:1972,Kolafa:1984,WangEtal:1989,WangEtal:1990,ParkWidom:1989,%
FerreiraSokal:1995}, the three-state Potts 
antiferromagnet on the Kagom\'{e} lattice \cite{Baxter:1970}, coverings of 
the square lattice by dimers of one color \cite{FisherStephenson:1963}
or of two colors \cite{RaghavanEtal:1997}, and the 
four-coloring of the edges of the square lattice \cite{Read:1992}.

All known models of this type can be mapped, configuration by 
configuration, to configurations of integer-valued `heights'
$\vecz(\vecx)$ \cite{vanBeijeren:1977,BloteHilhorst:1982,Kolafa:1984,%
ZhengSachdev:1989,Levitov:1990,HuseRutenberg:1992,KondevHenley:1995b,%
KondevHenley:1996}; we will call models with this property `height models'.
We can interpret $\vecz(\vecx)$ as parametrizing a surface, which
(in most of the above cases) turns out to be 
in its {\em rough} phase, 
described by a gradient-squared free energy at long wavelengths.  
By standard methods,
this corresponds to algebraically decaying spin-spin correlations.  
(See \eg{}  Nelson~\shortcite{Nelson:1983} and 
Nienhuis~\shortcite{Nienhuis:1987}; we 
review the derivation in Sec.~\ref{sec:2}). 
The exponents may be calculated from 
the stiffness constant in the long-wavelength Gaussian free energy, 
and indeed the most accurate way to determine the exponents is by 
directly determining this stiffness constant from the mean-square 
fluctuations of the Fourier transform of $\vecz(\vecx)$, as measured in Monte 
Carlo simulations \cite{KondevHenley:1995b,RaghavanEtal:1997,%
ZengHenley:1997}.
It can be seen that the spirit of this `height model' approach is geometrical:
spin fluctuations are visualized as undulations of an interface model, 
and critical exponents are obtained through computations of 
reciprocal lattice vectors in a high dimensional lattice. 

The height model we introduce here is a new way of generalizing the three-state 
Potts antiferromagnet (AFM) on the square lattice to general $q$ (and 
particularly to $q = 4$).  Recall that the three-state Potts AFM 
is in a ground state if and only if the four 
spins around every plaquette are (ABAB), (ABCB), or a configuration 
related to these by symmetry.  The usual $q$-state Potts AFM with $q>3$ allows 
an additional kind of plaquette configuration in the ground state, 
of type (ABCD), and it turns out that this excludes any height 
representation, which was responsible for the interesting critical 
properties of the $q = 3$ case.  (Indeed, the Potts AFM with $q>3$ has 
exponential rather than power-law decay of correlations 
\cite{GrestBanavar:1981}.)
Therefore, our model excludes the additional kind of plaquette so as 
to keep the height representation.
We call such a model a `constrained Potts 
antiferromagnet' (CPAFM).  Our motivation to study this model
is that the height space (in which 
the $\vecz$ vectors live) has an especially high dimension (five), 
and hence might exhibit a new universality class. 

In the remainder of this paper, we present a study of the 4-state 
CPAFM model on the square lattice.  First, in Sec.~\ref{sec:2}, we review 
the basic ideas of height models (using the 3-state Potts AFM as a 
specific example).  In particular, we explain how to express a 
spin operator in terms of the height correlation function.  In 
Sec.~\ref{sec:3} we carry this out for the more complicated case of the 
4-state CPAFM, presenting the height mapping of spin 
configurations and the height representation of spin operators.  
Sec.~\ref{sec:4} describes the cluster Monte Carlo algorithm needed for 
studying such models.  We present the simulation results, and 
discuss their implications for whether the system is in a critical 
or a long-range-ordered state, in Sec.~\ref{sec:5}.  Finally, Sec.~\ref{sec:6} 
summarizes our results in the context of related models.

In the present paper, we have considered only the zero temperature behavior. 
At small nonzero temperatures, there are rare
plaquettes where the constraint rule is violated. 
Each such point corresponds to a vortex-like defect (a dislocation)
in $\vecz(\vecx)$. In other height models, 
such defects are associated with further interesting exponents 
\cite{KondevHenley:1996}. Formulas for these exponents 
(assuming a rough surface)
would be straightforward to derive for the CPAFM model 
as an extension of  Sec.~\ref{sec:3}. 

%% file: CPAFM2.tex
\section{Example height model: The 3-state Potts AFM} \label{sec:2}

This section is a pedagogical introduction to some basic concepts
and methods which are valid for all ``height models''. 
Our of these, we chose the three-state Potts antiferromagnet (AFM)
on the square lattice (see Figure \ref{fig:3stateGroundState})
as the example model for its obvious analogies 
to the more elaborate 4-state model of Sec.~\ref{sec:3}.
(These analogies also led us to select it for numerical comparisons). 
Furthermore, this model
had never been explicitly analyzed using the geometrical 
``height-model'' language~\footnote{
The triangular Ising antiferromagnet {\it was} so analyzed
by Nienhuis \etal{}~\shortcite{NienhuisEtal:1984}.}, except
in the poorly-known simulation paper of
Kolafa~\shortcite{Kolafa:1984}.

\begin{figure}\tt
\settowidth{\figboxwidth}{A C A B A B A B$|$C B C B A B A B}
\addtolength{\figboxwidth}{25mm}
\parbox[t]{\figboxwidth}{(a)\\
\\
A C A B A B A B$|$C B C B A B A B\\
B A B A C A B A$|$B A B C B C B C\\
A C A C A C A B$|$A B C A C A C A\\
C A B A C A B A$|$B A B C A B A C\\
A B A B A C A C$|$A B C A B C B A\\
C A B A B A B A$|$B A B C A B A C\\
A B A C A C A B$|$C B A B C A C B\\
B A C A C A C A$|$B A C A B C B A}
\parbox[t]{\figboxwidth}{(b)\\
\\
2 1 2 3 2 3 2 3$|$4 3 4 3 2 3 2 3\\
3 2 3 2 1 2 3 2$|$3 2 3 4 3 4 3 4\\
2 1 2 1 2 1 2 3$|$2 3 4 5 4 5 4 5\\
1 2 3 2 1 2 3 2$|$3 2 3 4 5 6 5 4\\
2 3 2 3 2 1 2 1$|$2 3 4 5 6 7 6 5\\
1 2 3 2 3 2 3 2$|$3 2 3 4 5 6 5 4\\
2 3 2 1 2 1 2 3$|$4 3 2 3 4 5 4 3\\
3 2 1 2 1 2 1 2$|$3 2 1 2 3 4 3 2}
	\sepCaption{(a) A ground state of the 3-state Potts AFM.  On the 
	left side is a portion of the A B/C ideal state; in the middle, a 
	portion of the AB state (not an ideal state); on the right, a random 
	state.  (b) The corresponding height configuration.  (Note that the 
	heights are well-defined only after the height at one site is fixed.)}
	\label{fig:3stateGroundState}
\end{figure}

As already noted, a 
configuration is a ground state if and only if each plaquette is 
assigned either two or three distinct spins, with like spins situated 
on opposite corners of the plaquette.
Our object is to compute the critical exponent governing the decay 
of the correlation function of various operators.  The steps to 
this are, first, to map the spins to a height representation 
described by a gradient-squared free energy, which allows us to 
compute correlations of the height variables
(subsec.~\ref{sec:3stateHeightMap} below, 
in particular Eq.~\eqref{oneDHeightCorrelation});
second, to relate the spin operators to the 
height variables (subsec.~\ref{sec:3stateIdealStates} and 
\ref{sec:3stateExponents} below).  The same steps 
would be used for any other height model \cite{KondevHenley:1995a,%
KondevHenley:1995b,KondevHenley:1996,RaghavanEtal:1997} and in
particular for the 4-state 
CPAFM (in Sec.~\ref{sec:3}, below).~\footnote{The height map
described in subsec.~\ref{sec:3stateHeightMap} also
works for the $q=3$ Potts AFM on a simple cubic lattice.  
That model has long-range order in the ground state ensemble 
\cite{BanavarEtal:1980} as would be expected from the three-dimensional 
version of our arguments below.  At higher temperatures it almost exhibits
continuous symmetry behavior \cite{GottlobHasenbusch:1994,%
KolesikSuzuki:1995}. It is intriguing, then, to speculate whether its
coarse-grained height variable becomes the angle of an XY-model, as 
does happen in the two-dimensional case we describe in this section.}

In view of the application to
a multi-dimensional height variable in Sec.~\ref{sec:3}, 
the most important new concepts here are the three lattices 
associated with height space, defined in Subsec.~ \ref {sec:3stateOperators}.
In addition, the concept of 
`ideal state' is introduced in Sec.~\ref{sec:3stateIdealStates}
which eases (and geometrizes) 
the determination of the $G$ vector of a given operator.

The {\it results} of this section were already derived by 
den Nijs \etal{}~\shortcite{denNijsEtal:1982}, 
but those authors had a different focus: they downplayed the exact mapping to
height variables, while emphasizing the application of results from
renormalization group theory and from exact solutions of the eight-vertex
model. (For our results, a more elementary theoretical level will suffice.)
However, once the height-space reciprocal lattice vector 
$G$ is identified for a particular operator, 
the scaling indices $\eta(G)$ are found here 
(see Sec.~\ref{sec:3stateExponents})
exactly as in 
den Nijs \etal{}~\shortcite{denNijsEtal:1982}.

\subsection{Height mapping} \label{sec:3stateHeightMap}

For each ground state configuration, we define a height function 
$z(\vecx)$ such that the difference in height, $\Delta{}z$, in going from 
one site to a neighboring site is $+1$ if the pair of spins are in 
the same order as in A$\to$B$\to$C$\to$A and is $-1$ if they are in the 
reverse order.  In this way we have a height $z(\vecx)$ uniquely 
assigned to each site $\vecx$ once the height at any particular site is 
specified (see Figure~\ref{fig:3stateGroundState}(b)).
This $z(\vecx)$ is well-defined (since the 
sum of $\Delta{}z$ around any closed loop is zero) for any ground 
state.

We can coarse-grain the microscopic $z(\vecx)$ to obtain the macroscopic 
height $h(\vecx)$.  We now postulate an effective elastic free energy of 
the form
\begin{equation} \label{oneDFreeEnergy}
F = \int\dvecx\frac{1}{2}K |\del{}h(\vecx)|^{2},
\end{equation}
where K is a stiffness constant.  Writing \eqref {oneDFreeEnergy}
in terms of $\htilde(\vecq)$, the Fourier transform of $h(\vecx)$, gives
\begin{equation}
F = \sum_{\vecq}\frac{1}{2}K |\vecq|^{2} |\htilde(\vecq)|^{2},
\end{equation}
where the Fourier transform is normalized as
\begin{equation}
h(\vecx) = \frac{1}{\sqrt{N}} \sum_{\vecq}
						\e^{i \vecq \cdot \vecx}\htilde(\vecq)
\end{equation}
and $N$ is the number of sites in the lattice.  All sums over $q$ run 
over the first Brillouin zone.

Hence, the partition function is a product of independent 
Gaussians and the fluctuations in $\htilde(\vecq)$ are given by
\begin{equation} \label{oneDFluctuations}
\langle |\htilde(\vecq)|^{2} \rangle = \frac{1}{K |\vecq|^{2}}.
\end{equation}
Fourier transforming back to real space shows that the height 
correlation function is
\begin{equation} \label{oneDHeightCorrelation}
         \langle |h(\vecx')-h(\vecx)|^{2} \rangle
		\approx \frac{1}{\pi{}K} \log{r} + C,
\end{equation}
where $r \equiv |\vecr| \equiv |\vecx'-\vecx|$ is assumed to be large 
and $C$ is a constant.  Models for which 
$\langle |h(\vecx')-h(\vecx)|^{2} \rangle$
diverges as 
$r\to\infty$ (as in \eqref{oneDHeightCorrelation}) are termed 
``rough.''  Therefore, models for which \eqref{oneDFreeEnergy} is 
valid (with finite, nonzero stiffness $K$) are rough, as is the 
3-state Potts AFM.

\subsection{Ideal states} \label{sec:3stateIdealStates}
The first step in relating spin operators to heights is to define a 
set of reference states, called `ideal states', such that the spin, 
$\sigma(\vecx)$, on site $\vecx$ is a function of the coarse-grained 
height at $\vecx$.  Ideal states are configurations which (i) consist 
of a periodic pattern of spins, (ii) have zero macroscopic height 
gradient $\del{}h$ (and are therefore called `flat'), and (iii) have 
maximal entropy for local fluctuations.  
In general, ``ideal states'' make it convenient to identify
the height-space reciprocal lattice vector 
(see Subsec.~\ref{sec:3stateExponents}, below)
which is associated with the locking operator and with spin operators 
of interest
(see Subsec.~\ref{sec:3stateOperators}). 

The ideal states for the 
3-state Potts AFM are the A~B/C state (left part of 
Figure~\ref{fig:3stateGroundState}(a)) and the others equivalent to 
it by symmetry; in our nomenclature the labels refer to the even and 
odd sublattices, respectively, and a label `B/C' indicates a 
disordered site with spin having equal probability of B or C.  A first 
guess for an ideal state might have been the AB state (center part of 
Figure~\ref{fig:3stateGroundState}(a)), which is clearly periodic and 
flat.  However, the number of states `close to' the AB state is less 
than the number of states close to the typical A~B/C state.  (Here 
distance between states is measured by the number of spin flips 
required to convert one into the other.)  Hence, the AB state is not 
an ideal state, but rather a transition state between A~B/C and C/A~B 
states.

All the sites in the completely ordered sublattice of an ideal 
state (\eg{} the even sites in an A~B/C state) clearly have the same 
microscopic height $z=z_{\nought}$.  The values of $z$ on the sites in the 
disordered lattice are $z_{\nought} \pm 1$, with the sign depending on the 
particular value of the spin on that site.  Since a disordered site 
is equally likely to have either of the two possible spins, it is 
clear that the macroscopic height of the ideal state is $h=z_{\nought}$.

We expect a typical ground state configuration to consist primarily 
of macroscopic domains of ideal states, separated by thin walls of 
transition states.  The value of $h$ within a domain is simply the 
value of $z$ on the disordered sublattice of the domain.  Therefore, 
the macroscopic height difference, $\Delta{}h$, between two ideal 
domains can be calculated by finding the net height change along a 
path of neighboring sites connecting a site $\vecx$ in the ordered 
sublattice of the first domain with a site $\vecx'$ in the ordered 
sublattice of the second domain:
\begin{equation} \label{oneDDeltaH}
\Delta{}h = \sum_{i=0}^{n-1} \Delta{}z(\sigma(\vecx_{i}),\sigma(\vecx_{i+1}))
			= \sum_{i=0}^{n-1} \Delta{}z(\sigma_{i},\sigma_{i+1}),
\end{equation}
where $n$ is the number of bonds in the chosen path connecting $\vecx$ 
to $\vecx'$, $\vecx_{i}$ and $\vecx_{i+1}$ are nearest neighbors for 
$i=0$ to $n-1$, $\sigma_{i}=\sigma(\vecx_{i})$ for $i=0$ to $n$, 
$\vecx_{0} = \vecx$, and $\vecx_{n} = \vecx'$.  Note that $n$ is even 
if the same sublattice is ordered in both domains, and $n$ is odd if 
opposite sublattices are ordered in the two domains.  It is not 
difficult to show (see Appendix~\ref{sec:appA}) that the macroscopic height 
difference between two domains is uniquely determined (modulo $6$) by 
the identities of the ideal states in the two domains.  Whether two 
patches of (say) A~B/C domain have the same height, or heights which 
differ by 
$\pm 6$ (or more), is not arbitrary but depends on the configurations 
between the domains.  For example, if we follow a path through a 
succession of domains A~B/C, C/A~B, C~A/B, B/C~A, A/B~C, and back to 
A~B/C, each has a macroscopic height $h$ larger by $1$ than the previous 
domain, so that the second A~B/C domain has a height that is $6$ 
greater than the first A~B/C domain.
(See Figure~\ref{fig:3stateIdealStates}(a).)
\begin{figure}
\epsfig{file=fig02.epsi,angle=+90,width=5.5in}
	\sepCaption{(a) The six ideal states of the 3-state Potts AFM 
	arranged in a circle, with heights as indicated.
	(b) The `ideal-states' graph, with repeat distance $6$, and  
	operator components $\stagMagMoment$ and $\magMoment$ shown as a 
	function of $h$.  The operator $P$ (not shown) would alternately 
	take values $+\frac{1}{2}$ and $-\frac{1}{2}$.}
	\label{fig:3stateIdealStates}
\end{figure}
Once the microscopic height on a particular site is set, the ideal 
state pattern (and in particular the spin value on the ordered 
sublattice) within every ideal-state domain can be identified 
uniquely from its height $h$.

\subsection{Correlation exponents} \label{sec:3stateExponents}
Our goal is to compute the correlation function of a given local
operator $\opO(\vecx)$, which is some function of the spins
$\sigma(\vecx')$ for $\vecx'$ in a neighborhood of $\vecx$. 
To accomplish this, we must first convert this operator from
a function of the local spin variables to a function of the 
local height variables. 
Namely, each site belongs to a domain of some ideal state;
the spin variable on that site is determined by the ideal state label 
of that domain, 
and (as emphasized in Subsec.~\ref{sec:3stateIdealStates})
every ideal state  is  labeled unambiguously by its height $h$. 

Since the periodicity of the spin arrangement in each ideal state 
divides the lattice into `even' and `odd' sublattices, any local 
operator $\opO(\vecx)$ will have similar periodicity and dependence 
on which sublattice $\vecx$ is in:
\begin{equation} \label{oneDOperator}
\opO(\vecx) = F_{\opO}^{\parity(\vecx)}(h),
\end{equation}
where $\parity(\vecx)$ is the parity of site $\vecx$ 
($\parity(\vecx)=1$ if $\vecx$ is in the even sublattice and 
$\parity(\vecx)=-1$ if $\vecx$ is in the odd sublattice).
In addition, since $\opO(\vecx)$ depends on $h$ only through its 
dependence on the ideal state label, we have
\begin{equation} \label{oneDPeriodicity}
F_{\opO}^{\parity(\vecx)}(h) = F_{\opO}^{\parity(\vecx)}(h+\aperp),
\end{equation}
where $\aperp$ is any integer multiple of $6$.

In light of \eqref{oneDOperator} and \eqref{oneDPeriodicity} the 
function $F_{\opO}^{\parity(\vecx)}$ can be expanded as a Fourier 
series:
\begin{equation} \label{oneDFourier}
\opO(\vecx) = F_{\opO}^{\parity(\vecx)}(h) =
	\sum_{G} \opO_{G}^{\parity(\vecx)} \e^{\iunit{}Gh},
\end{equation}
where $G$ ranges over the integer multiples of $2\pi/6$.  From 
\eqref{oneDFourier} we conclude that the correlations of 
$\opO(\vecx)$ satisfy
\begin{equation} \label{oneDOpAverage}
\langle \opO(\vecx) \opO(\vecx') \rangle =
	\sum_{G,G'} \opO_{G}^{\parity(\vecx)}
	\conjugate{\opO_{G'}^{\parity(\vecx')}}
		\langle \e^{\iunit(Gh-G'h')} \rangle,
\end{equation}
where $h$ is the height at $\vecx$, $h'$ is the height at $\vecx'$, 
and $\langle \ldots \rangle$ denotes the average over allowed 
configurations.  Note that
\begin{equation} \label{oneDExpAverages}
\langle \e^{\iunit{}(Gh-G'h')} \rangle =
	\langle \e^{\iunit{}G(h-h')}\e^{\iunit{}(G-G')h'} \rangle =
	\langle \e^{\iunit{}G(h-h')} \rangle \langle \e^{\iunit{}(G-G')h'} \rangle,
\end{equation}
where the second equality follows because $(h-h')$ is independent of 
$h'$.  But $\langle \e^{\iunit{}(G-G')h'} \rangle = \delta_{G,G'}$, so 
that substituting \eqref{oneDExpAverages} into \eqref{oneDOpAverage} 
gives
\begin{equation} \label{oneDOpAvg2}
\langle \opO(\vecx) \opO(\vecx') \rangle =
	\sum_{G} \opO_{G}^{\parity(\vecx)} \conjugate{\opO_{G}^{\parity(\vecx')}} 
		\langle \e^{\iunit{}G(h-h')} \rangle.
\end{equation}
Since $\{\htilde(\vecq)\}$ are Gaussian distributed, so is $h(\vecx) - 
h(\vecx')$.  Moreover, $\langle h(\vecx) - h(\vecx') \rangle$ is 
clearly zero.  Hence, we can use \eqref{oneDHeightCorrelation} to show
\begin{equation} \label{oneDAlgDecay}
\langle \e^{\iunit{}G(h-h')} \rangle =
	\e^{-\frac{1}{2}G^{2}\langle |h-h'|^{2} \rangle} \approx
	c(|G|) r^{-\eta_{G}} \text{ as }r \to \infty,
\end{equation}
where
\begin{equation} \label{oneDEtaG}
\eta_{G} = \frac{G^{2}}{2\pi{}K}
\end{equation}
and $c(|G|) = \e^{-G^{2}C/2} > 0$.  With \eqref{oneDAlgDecay}, 
equation~\eqref{oneDOpAvg2} becomes
\begin{equation} \label{oneDOpCorr}
\begin{split}
\langle \opO(\vecx) \opO(\vecx') \rangle &\approx
	\sum_{G} c(|G|) \opO_{G}^{\parity(\vecx)} 
		\conjugate{\opO_{G}^{\parity(\vecx')}} r^{-\eta_{G}}\\
	&\approx f(\parity(\vecx),\parity(\vecx')) r^{-\eta},
\end{split}
\end{equation}
where
\begin{equation} \label{oneDEta}
\eta = \min \left\{\eta_{G}: G = \frac{2\pi{}n}{6};
		n\in\Naturals; \opO_{G}^{+1}\text{ or }
		\opO_{G}^{-1}\text{ is nonzero}\right\}
	\equiv \eta_{G_{\opO}},
\end{equation}
$c = c(G_{\opO})$, and $f(\parity,\parity') = 2c \re
\left\{\opO_{G_{\opO}}^{\parity} 
\conjugate{\opO_{G_{\opO}}^{\parity'}}\right\}$.
Note that $G_{\opO}$ is the smallest positive wavevector where the 
(height space) Fourier transform is nonzero.  Also, $f(\parity,\parity')$ 
is not identically zero if and only if at least one of $\opO_{G}^{+1}$ 
and $\opO_{G}^{-1}$ is nonzero -- and this is guaranteed by 
\eqref{oneDEta}.  Strictly speaking, equation~\eqref{oneDOpCorr} is 
valid as written only if $f(\parity,\parity')$ is nonzero for all 
four possible argument combinations.  If this is not true, then for 
$(\vecx,\vecx')$ where $f(\parity(\vecx),\parity(\vecx')) = 0$ we 
should interpret \eqref{oneDOpCorr} as saying
$\langle \opO(\vecx) \opO(\vecx') \rangle \ll r^{-\eta}$.

In any case, we see that an effective elastic free energy of the 
form \eqref{oneDFreeEnergy} implies that correlations in any 
(nontrivial) operator decay algebraically -- \ie{}, there is 
quasi-long-range order.  Moreover, once the stiffness $K$ is known, 
the decay exponent $\eta$ for correlations in any operator can be 
determined from \eqref{oneDEtaG} and \eqref{oneDEta}.  Thus, all correlation 
exponents are known once any of them is determined.

\subsection{Specific operators} \label{sec:3stateOperators}
Now we are ready to predict (in terms of the stiffness $K$) the 
exponents for explicit operators in the 3-state Potts AFM model.  
To do this, one should (for a given site $\vecx$) make a plot of
the value of $O(\vecx)$ as a function of the coarse-grained height $h$
that labels the ideal state, as shown in Fig. 2; the Fourier components
(as in Eq. 9) can then be read off, along with the coefficients
$\opO_{G}^{\parity(\vecx)}$ (which are not usually needed). 

In order to define the operators, 
it is helpful to represent the spin state on each site by a unit vector 
$\magMoment(\vecx)$ in the plane, pointing at angles $0$, $2\pi/3$, 
and $4\pi/3$, for spin states A, B, and C, respectively.
We must first identify the value of $G_{\rm O}$ corresponding to a given
operator. 

The {\em staggered} magnetization operator $\stagMagMoment(\vecx) 
\equiv \parity(\vecx) \magMoment(\vecx)$ is the order parameter since 
it distinguishes among the $6$ ideal states and also has the maximum 
height-space period of $6$, as shown in
Figure~\ref{fig:3stateIdealStates}(b).  Thus, the minimizing 
wavevectors in \eqref{oneDEta} are $G_{S} = \pm \dfrac{2\pi}{6}$.  The value 
$\eta_{S} = \dfrac{\pi}{18K}$ then follows directly from 
\eqref{oneDEtaG}.  Wang \etal{}~\shortcite{WangEtal:1989} found 
$\eta_{S}$ numerically from Monte Carlo simulations.  The exponents 
were obtained analytically by Park and 
Widom~\shortcite{ParkWidom:1989}, using the exact mapping of the 
3-state Potts AFM to the 6-vertex model and the known exact solution 
of the latter.  They found the free energy cost of `step' boundary 
conditions, forcing $\Delta{}h = \pm 2$ across the system, to be 
$\Delta{}F = 2\pi/6L$; via conformal invariance this implied 
$\eta_{S} = 1/3$.  Given that this is a height model, one could 
instead directly extract the stiffness $K = \pi/6$ by inserting the 
forced tilt $\Delta{}h/L$ into \eqref{oneDFreeEnergy} (which gives 
$F = K(\Delta{}h)^{2}/2L$).

The ideal states are `ferrimagnetic' -- \ie{}, they each have a net 
magnetization $\magnetization$; this operator has a height-space 
period of $3$ (see Figure~\ref{fig:3stateIdealStates}(b)).  
Therefore, $G_{M} = \pm 2\pi/3 = 2G_{S}$ and $\eta_{M} = 4\eta_{S} = 
4/3$.  The ferromagnetic exponent $\eta_{M}$ seems to be considered 
explicitly only by den Nijs \etal{}~\shortcite{denNijsEtal:1982}, who 
used a Coulomb-gas approach very similar to ours to reach the same 
conclusions.
This exponent is relevant and could easily be measured in 
simulations, but this has apparently never been implemented.

An ideal state can have either the even or odd sublattice as the 
ordered one.  These possibilities can be distinguished by the sign of 
the operator $\displaystyle P(\vecx) = \frac{1}{4} \parity(\vecx) 
\sum_{\text{n.n.n. }\vecx'} \delta_{\sigma(\vecx),\sigma(\vecx')}$, 
where the sum is over $\vecx'$ which are second neighbors with $\vecx$.
The height-space period of this operator is clearly $2$ and, hence, 
$G_{sub} = \pm 2\pi/2$.  The exponent for this operator is irrelevant 
for the (ground state) 3-state Potts AFM model.

At this point we may identify three one-dimensional lattices which 
appear in the height representation of the three-state Potts AFM. 
First, the ``repeat lattice'' in height space has period 6, because
(as noted in Sec. 2.2) the spin pattern as a function of $h$ repeats 
itself with that period. 
Secondly, the ``equivalence lattice'' has period 1, since 
$h \to h+1$ is induced by a global permutation of Potts spins to give
a symmetry-equivalent state. 
Finally, the ``height space reciprocal lattice'' is dual to the
repeat lattice and thus has period $2\pi/6$ for this model;
it is a general fact about height models that
the allowed $G$ values (see Sec. 2.3) belong to the reciprocal lattice. 
These lattices were not defined in prior applications of height models, 
since they are rather trivial in one dimension; they begin to become
useful when one wants to make a systematic comparison between different
height models, and they are quite essential in dealing with 
multi-dimensional height models, as in the next section.

%% file: CPAFM3.tex
\section{The 4-state constrained Potts AFM} \label{sec:3}
A more complicated model with a height mapping is the 4-state 
constrained Potts antiferromagnetic (CPAFM) model on the square 
lattice (see Figure~\ref{fig:4stateGroundState}(a)).
\begin{figure}\tt
\settowidth{\figboxwidth}{A C A D$|$A B C A}
\addtolength{\figboxwidth}{10mm}
\parbox[t]{\figboxwidth}{(a)\\
\\
A C A D$|$A B C A\\
B A B A$|$C A B C\\
A C A B$|$A C A B\\
C A D A$|$D A D A\\
A B A C$|$A B A C\\
D A D A$|$B C B A\\
A C A B$|$D B D B\\
B A C A$|$B A B C}
\parbox[t]{\figboxwidth}{(b)\\
\\
2 3 2 1$|$2 2 1 0\\
2 2 2 2$|$3 2 2 1\\
2 3 2 2$|$2 3 2 2\\
3 2 1 2$|$1 2 1 2\\
2 2 2 3$|$2 2 2 3\\
1 2 1 2$|$2 1 2 2\\
2 3 2 2$|$3 2 3 2\\
2 2 3 2$|$2 2 2 1}
\parbox[t]{\figboxwidth}{(c)\\
\\
2 1 2 1$|$2 4 5 4\\
4 2 4 2$|$1 2 4 5\\
2 1 2 4$|$2 1 2 4\\
1 2 1 2$|$1 2 1 2\\
2 4 2 1$|$2 4 2 1\\
1 2 1 2$|$4 5 4 2\\
2 1 2 4$|$5 4 5 4\\
4 2 1 2$|$4 2 4 5}
	\sepCaption{(a) Portion of a ground state of the 4-state CPAFM.  The 
	left portion shows the A~B/C/D ideal state, while the right portion 
	(separated by a vertical dashed line) shows a random state.
	(b) The corresponding configuration of even height component $h_{1}$.
	(c) The corresponding configuration of odd height component $h_{4}$.}
	\label{fig:4stateGroundState}
\end{figure}
Now there are four possible spins -- A, B, C, and D -- and, on top of 
the antiferromagnetic nearest-neighbor constraint, there is the added 
constraint that no plaquette may contain more than $3$ distinct spin 
states.  In particular, as in the 3-state Potts AFM model, a 
configuration is a ground state if and only if each plaquette is 
assigned either two or three distinct spins, with like spins situated 
on opposite corners of the plaquette.

In this section, we will work through all the stages of analyzing this
model via its height representation. These are 
(i) discovering all
vector components of the height representation; 
(ii) determining the generic elastic theory;
(iii) mapping the ideal-states as a point set in height space
(associated with this is the ``repeat lattice'');
(iv) evaluating the height-space reciprocal lattice vectors, 
and finally the exponents, corresponding to the operators of interest. 

\subsection{Height mapping and elastic theory}
\label{sec:4stateHeightMap}
Unlike in the 3-state case, height space for the 4-state CPAFM model 
is multidimensional.  It has dimension $\dperp = 5$ and can be viewed 
as the Cartesian product of a 3-dimensional `even' height subspace and a 
2-dimensional `odd' height subspace.  For these we shall sometimes use 
the notation $\evenv{a,b,c} \equiv (a,b,c,0,0)$ and $\oddv{a,b} 
\equiv (0,0,0,a,b)$.)

The change in height in going from a site $\vecx$ to a neighboring 
site $\vecx'$ is $\vecz(\vecx') - \vecz(\vecx) = \Delta\vecz =
\DzE \otimes \DzO$, where 
$\DzE \equiv (\Delta{}z_{1}, \Delta{}z_{2}, \Delta{}z_{3})$ and
$\DzO \equiv (\Delta{}z_{4}, \Delta{}z_{5})$ are the even and odd 
height change functions, respectively.  Here $\DzE = 
\DzE(\sigma(\vecx), \sigma(\vecx'))$ is a function only of the spins 
on the two sites and does not depend on which site belongs to which 
sublattice.  On the other hand, $\DzO = \DzO(\parity(\vecx), 
\sigma(\vecx), \sigma(\vecx'))$ depends also on the parity 
$\parity(\vecx)$ of the initial site but does not depend on which spin 
is on which site, \ie{} $\DzO(\parity, \sigma, \sigma') = 
\DzO(\parity, \sigma', \sigma)$.

The identification of the correct height space is nontrivial:
for example, had we simply generalized the $q=3$ case, 
we would have omitted the ``odd'' components, 
since the 3-state height variable is ``even''. 
On the other hand, there is also a danger of including
redundant height components. By redundant, we mean functions of the spins which 
are linear combinations of other height variables and of {\em bounded} 
functions of the spins. 
Appendix~\ref{sec:app0} gives the explicit construction of 
$\vecz(\vecx)$ confirming the validity of its definition. 

All values of $\Delta\vecz$ can be found from Table~\ref{table:heightSteps}, which 
directly shows the value of $\DzE(\sigma,\sigma')$ for $12$ of the 
$24$ possible argument combinations and the value of 
$\DzO(\parity,\sigma,\sigma')$ for $24$ of the $48$ possible argument 
combinations.
Values of $\Delta\vecz$ for the remaining argument combinations 
follow from $\DzE(\sigma,\sigma') = -\DzE(\sigma',\sigma)$ and 
$\DzO(\parity,\sigma,\sigma') = -\DzO(-\parity,\sigma',\sigma)$, which 
must hold for $\Delta\vecz(\vecx)$ to be single-valued.  
Figure~\ref{fig:4stateGroundState}(b,c) shows an even and an odd 
component of height for the example configuration.

The primitive vectors for the `odd' subspace, $\oddv{1,0}$ and
$\oddv{0,1}$, have been chosen to have a $120^{\circ}$ angle between them
as is standard for a triangular lattice; thus the length of $\oddv{x_4,x_5}$
is $(x_4^2+x_5^2-x_4x_5)^{1/2}$. 
\footnote {The symmetry in the definition of the `odd' 
heights would be more evident using an alternative representation using three
`odd' components with the constraint $z_{4} + z_{5} + z_{6}=0$.}
The {\it reciprocal} space `odd' vectors, though, are defined so that
(height) reciprocal-space $\oddv{1,0}$ 
has a dot product 
with height-space $\oddv{1,0}$  of unity, 
and with height-space $\oddv{0,1}$  of zero;
and such that reciprocal space $\oddv{1,0}$ and $\oddv{0,1}$ are 
$60^{\circ}$ apart, so $\oddv{q_4,q_5}$ has length 
$(q_4^2+q_5^2+q_4q_5)^{1/2}$. 

\begin{table}
	\begin{tabular}{|l|>{$}c<{$}|}
		\hline
		Spin pair & \Delta\vecz \\ \hline
		AB & \evenv{0,1,-1} + \parity\oddv{2,-1} \\ \hline
		AC & \evenv{1,0,1} + \parity\oddv{-1,2} \\ \hline
		AD & \evenv{-1,-1,0} + \parity\oddv{-1,-1} \\ \hline
		BC & \evenv{-1,1,0} + \parity\oddv{-1,-1} \\ \hline
		BD & \evenv{1,0,-1} + \parity\oddv{-1,2} \\ \hline
		CD & \evenv{0,1,1} + \parity\oddv{2,-1} \\ \hline
	\end{tabular}
	\sepCaption{Rules for height steps.}
	\label{table:heightSteps}
\end{table}
As with the 3-state model, $\vecz(\vecx)$ is well-defined for any 
allowed spin configuration once its value at any particular site is 
specified, and we can coarse-grain the microscopic $\vecz(\vecx)$ to 
obtain the macroscopic height $\vech(\vecx)$.  The higher-dimensional 
analog of the elastic free energy in \eqref{oneDFreeEnergy} is
\begin{equation} \label{Kcontraction}
      F = \int \dvecx \frac{1}{2} 
      \sum_{\alpha,\beta=1}^{5} 
	K_{\alpha\beta}\del{}h_{\alpha} \cdot \del{}h_{\beta}.
\end{equation}
A detailed consideration of the symmetries of the square lattice 
was needed to verify that the coefficients 
in \eqref{Kcontraction}
are diagonal in the gradient indices.
(In a general height model, they 
might depend on {\it four} indices, 
two for the gradients in addition to the two for height components.)
The symmetries of the Potts spin further restrict the matrix $\tensorK = 
\left(K_{\alpha\beta}\right)$  to have the form
\begin{equation} \label{Kmatrix}
\tensorK =
\begin{pmatrix}
	\KE & 0 & 0 & 0 & 0 \\
	0 & \KE & 0 & 0 & 0 \\
	0 & 0 & \KE & 0 & 0 \\
	0 & 0 & 0 & \KO & -\frac{1}{2}\KO \\
	0 & 0 & 0 & -\frac{1}{2}\KO & \KO
\end{pmatrix},
\end{equation}
where the even and odd height subspace stiffness constants, $\KE$ 
and $\KO$, are as yet undetermined.  (The `odd' part of
the density in \eqref{Kcontraction} is just
$\frac{1}{2} \KO |\del\vech^{\Oscript}|^{2}$; the corresponding subblock in 
\eqref{Kmatrix} looks nondiagonal due to our choice of primitive 
vectors for the ``odd'' subspace.)
  The generalization of \eqref{oneDFluctuations} is
\begin{equation} \label{genFluctuations}
\langle \htilde_{\alpha}(\vecq) \conjugate{\htilde_{\beta}(\vecq)} 
	\rangle = \frac{\left(\tensorK^{-1}\right)_{\alpha\beta}}
					{|\vecq|^{2}}
\end{equation}
and, similarly, the analog of equation \eqref{oneDHeightCorrelation} 
is
\begin{equation} \label{genHeightCorrelation}
\langle \left[h_{\alpha}(\vecx') - h_{\alpha}(\vecx)\right]
		\left[h_{\beta}(\vecx') - h_{\beta}(\vecx)\right] \rangle \approx
	\frac{\left(\tensorK^{-1}\right)_{\alpha\beta}}{\pi}\log(r) +
		C_{\alpha\beta},
\end{equation}
where $C_{\alpha\beta}$ is the $\alpha\beta$ component of a constant 
matric $\tensorC$.  The term `rough' is used to describe any height 
component, $h_{\alpha}$, for which $\langle
|h_{\alpha}(\vecx') - h_{\alpha}(\vecx)|^{2} \rangle \to \infty$ as 
$r \to \infty$.  Hence, the even (odd) height components in the 
4-state CPAFM model are rough if and only if $\KE$ ($\KO$) is finite.

\subsection{Ideal states}
The ideal states in the 4-state CPAFM model are like those discussed 
in Sec.~\ref{sec:3stateIdealStates} for the 3-state model, except with 
three possible spins on each site in the disordered sublattice instead 
of just two\footnote{This is like the intermediate-temperature state 
in the {\em unconstrained} 4-state Potts 
AFM~\cite{GrestBanavar:1981}; however, the low-temperature behavior 
of that model may be better described by two states disordered on the 
even sublattice and the other two spin states on the odd sublattice 
(compare Banavar \etal{}~\shortcite{BanavarEtal:1980}).}
(see Figure~\ref{fig:4stateGroundState}(a), left part); for example, 
the `A~B/C/D' state with A on the even sublattice and disordered B/C/D 
on the odd sublattice.  We shall sometimes use the abbreviated label `X' 
for the disordered lattice, thus this ideal state is designated `AX'.  
Since there are two choices for which sublattice is ordered and four 
choices for which spin appears on the ordered sublattice, there are 
$8$ ideal states.

All of our steps here are closely analogous to those for the 3-state 
model.  All the sites in the completely ordered sublattice of an ideal 
state have the same macroscopic height $\vecz = \vecz_{\nought}$.  The sites 
in the disordered lattice have height values $\vecz_{\nought} + 
\Delta\vecz$, where $\Delta\vecz$ has three possible values, 
corresponding to a step from the ordered spin to neighbors with the 
three possible kinds of spin state on the disordered sublattice.  
Since these $\Delta\vecz$ vectors add up to zero, and a disordered 
site is equally likely to have any of the three possible spin states, 
the macroscopic height of the ideal state is $\vech = \vecz_{\nought}$.

The macroscopic height difference between two ideal domains can be 
calculated by finding the height change along a path of neighboring 
sites connecting a site $\vecx$ in the ordered sublattice of the first 
domain to a site $\vecx'$ in the ordered sublattice of the second 
domain.  The generalization of equation~\eqref{oneDDeltaH} is simply
\begin{equation} \label{CPAFMDeltaH}
\Delta\vech = \sum_{i=0}^{n-1} 
	\Delta\vecz(\parity(\vecx_{i}), \sigma(\vecx_{i}), \sigma(\vecx_{i+1})) 
	= \sum_{i=0}^{n-1} \Delta\vecz(\parity_{i},\sigma_{i},\sigma_{i+1}).
\end{equation}
Note that now we must keep track of the site's parity $\parity(\vecx)$ 
in addition to the sequence of spins encountered.  Altering this 
parity leaves $\Delta\vechE$ unchanged while causing $\Delta\vechO$ to 
change sign.

The differences \eqref{CPAFMDeltaH} allow us to map out the
ideal-states graph for the 3-state CPAFM model, of which each 
node is the $\vech$ of the corresponding ideal state. 
Neighboring ideal states (those which differ by the minimum number of 
spin changes) such as AX and XB are related by a $\Delta\vech$ vector 
which is simply the $\Delta\vecz$ vector for the step AB, as listed in 
Table~\ref{table:heightSteps}, and similarly for all other pairs.  The 
ideal-state lattice is 3-co\"{o}rdinated (as is visible in either the 
`even' or `odd' projection) since, for any ideal state, there are 3 
ways to make another ideal state with the minimal change.  Ideal 
states related by exchanging even and odd sublattices turn out to be 
related by the shiftvectors in the upper half of
Table~\ref{table:shiftvectors}, which are simply computed from
Table~\ref{table:heightSteps}; the parity $\parity$ is that of the
first site in the string.

The `even' space projection of the ideal-states graph
is a `hydrogen peroxide lattice.'
(See Fig.~1 of Leu 
\etal{}~\shortcite{LeuEtal:1969}, and references therein.
The Bravais lattice of this projection is body-centered cubic (bcc) 
with lattice constant $4$.  The crystallographic basis is $4$ points 
per primitive bcc cell, at $[0,0,0]$ (assumed to represent the ideal 
state `AX') and at $[0,1,-1]$, $[1,0,1]$, and 
$[-1,-1,0]$.\footnote{The same lattice is also known as the `Laves 
graph of degree three'~\cite{Coxeter:1955} and is the backbone of the 
minimal surface known as the `gyroid' and found in bicontinuous 
systems~(S. Milner, personal communication).
It is interesting to compare this graph with Fig.~4 of Kondev and 
Henley~\shortcite{KondevHenley:1995b}, which is the three-dimensional
ideal-states graph for a different 4-coloring model.}
All these points are equivalent by symmetry, and the nearest-neighbor 
bonds make $120^{\circ}$ angles.  Analogously, the `odd' space 
projection is a honeycomb lattice with bonds of length $\sqrt{6}$.

\begin{table}
\begin{tabular}{|l|>{$}c<{$}|}
\hline
Spin sequence & \Delta\vecz \\ \hline
ABCA & \evenv{-2,2,-2} + \parity\oddv{2,2} \\ \hline
ABDA & \evenv{2,2,-2} + \parity\oddv{2,-4} \\ \hline
ACDA & \evenv{2,2,2} + \parity\oddv{-4,2} \\ \hline
ACBA & \evenv{2,-2,2} + \parity\oddv{2,2} \\ \hline
ADBA & \evenv{-2,-2,2} + \parity\oddv{2,-4} \\ \hline
ADCA & \evenv{-2,-2,-2} + \parity\oddv{-4,2} \\ \hline
\hline
ACBDA & \evenv{4,0,0} + \parity\oddv{0,6} \\ \hline
ABCDA & \evenv{0,4,0} + \parity\oddv{6,0} \\ \hline
ACDBA & \evenv{0,0,4} + \parity\oddv{-6,6} \\ \hline

ADBCA & \evenv{-4,0,0} + \parity\oddv{0,-6} \\ \hline
ADCBA & \evenv{0,-4,0} + \parity\oddv{-6,0} \\ \hline
ABDCA & \evenv{0,0,-4} + \parity\oddv{6,-6} \\ \hline
\end{tabular}
\sepCaption{Height differences for three-spin and four-spin
strings of spins (returning to the first spin state)}
\label{table:shiftvectors}
\label{table:shift4}
\end{table}

\subsection{Repeat lattice} \label{sec:4stateRepeatLattice}
Before considering operators as periodic functions of the height, we 
must understand the `repeat lattice'.  In general this is defined as 
the set of possible macroscopic height differences between domains of 
the {\em same} ideal state in configurations allowed by the model.  
(Note the distinction from the Bravais lattice of the ideal-states 
graph.)  Conversely, the macroscopic height difference $\Delta\vech$ 
between domains of {\em different} ideal states can never belong to 
the repeat lattice.  (This follows since $\Delta\vech$ can never be 
zero in such a case, which can be checked explicitly for our 
models.)  The repeat lattice of the 3-state Potts AFM is just the set 
of integer multiples of $6$.

It can be checked (see Appendix~\ref{sec:appA}) that the repeat 
lattice of the 4-state CPAFM is a centered hyper-tetragonal lattice in 
5-space generated by the vectors
\footnote{These happen to be twice the vectors in the lower half of
Table~\ref{table:shiftvectors}.}
\begin{equation} \label{repeatGen}
\left\{
\veca_{i}^{\pm} = \frac{1}{2}\left(\vecaE_{i} \pm \vecaO_{i}\right) :
	i \in \{1,2,3\}
\right\},
\end{equation}
where
\begin{equation}
\begin{gathered}
\vecaE_{1} = \evenv{8,0,0} \quad \vecaE_{2} = \evenv{0,8,0} \quad \vecaE_{3} = 
	\evenv{0,0,8}\\
\vecaO_{1} = \oddv{0,12} \quad \vecaO_{2} = \oddv{-12,0} \quad \vecaO_{3} = 
	\oddv{12,-12}.
\end{gathered}
\end{equation}
The vectors \eqref{repeatGen} span a space of only rank 5, as 
expected since $\dperp = 5$.  To represent the repeat lattice it is 
natural to choose the unit cell with $\vecaE_{1}$, $\vecaE_{2}$, 
$\vecaE_{3}$, $\vecaO_{1}$, and $\vecaO_{2}$ as edges along with the 
$8$-point basis $\left\{ A_{1}\veca_{1}^{+} + A_{2}\veca_{2}^{+} + 
A_{3}\veca_{3}^{+} : A_{1},A_{2},A_{3} \in \{0,1\} \right\}$.

\subsection{Correlation exponents} \label{sec:4stateExponents}
As in the 3-state Potts AFM, the periodicity of the spin arrangement 
in each ideal state divides the lattice into even and odd 
sublattices, so that any local operator $\opO(\vecx)$ has the form
\begin{equation}
\opO(\vecx) = F_{\opO}^{\parity(\vecx)}(\vech).
\end{equation}
Again, $\opO(\vecx)$ depends on height only in that it depends on the 
ideal state ensemble, so that
\begin{equation}
F_{\opO}^{\parity(\vecx)}(\vech) = F_{\opO}^{\parity(\vecx)}(\vech + 
		\vecaperp),
\end{equation}
where $\vecaperp$ is any vector in the repeat lattice.  We can 
therefore expand $F_{\opO}^{\parity(\vecx)}$ in a Fourier series to 
obtain
\begin{equation}
\opO(\vecx) = F_{\opO}^{\parity(\vecx)}(\vech) =
	\sum_{\vecG}\opO_{\vecG}^{\parity(\vecx)} \e^{\iunit\vecG\cdot\vech},
\end{equation}
where the sum ranges over all $\vecG$ in the repeat lattice's 
reciprocal lattice, which we henceforth call the 
`height-space reciprocal lattice.'
Proceeding as for the 3-state Potts AFM, but 
using \eqref{genHeightCorrelation} in place of 
\eqref{oneDHeightCorrelation}, shows that
\begin{equation} \label{genOpCorr}
\langle \opO(\vecx) \opO(\vecx') \rangle \approx 
F(\parity(\vecx),\parity(\vecx')) r^{-\eta} \text{ as }r \to \infty,
\end{equation}
where
\begin{gather}
\eta = \min\left\{ \eta_{\vecG} : \vecG \neq 0; \opO_{\vecG}^{+1} 
	\text{ or } \opO_{\vecG}^{-1} \text{ is nonzero} 
	\right\},\label{genEta}\\
\eta_{\vecG} = \frac{1}{2\pi} 
			\vecG\cdot\tensorK^{-1}\vecG,\label{genEtaG}
\end{gather}
$\displaystyle
F(\parity,\parity') = \sum_{\vecG\in{}G(\eta)} c(\vecG) \re\left\{ 
\opO_{\vecG}^{\parity} \conjugate{\opO_{\vecG}^{\parity'}} \right\}$, 
$c(\vecG) = \e^{-\vecG\cdot\tensorC\vecG/2}$, and $G(\eta) = 
\left\{ \vecG : \eta_{\vecG} = \eta \right\}$.
In each case $\vecG$ is in the repeat lattice's reciprocal lattice.  
Again, equation \eqref{genEta} guarantees that $F(\parity,\parity')$ 
is not identically zero, and equation \eqref{genOpCorr} should be read 
as $\langle \opO(\vecx) \opO(\vecx') \rangle \ll r^{-\eta}$ if it 
happens that $F(\parity(\vecx),\parity(\vecx')) = 0$.

We see from the above and equation \eqref{Kmatrix} that the 4-state 
CPAFM exhibits quasi-long-range order if and only if either $\KE$ 
or $\KO$ is finite.  If they are both finite, the algebraic decay 
exponent for the correlations in any local operator can be determined 
once $\KE$ and $\KO$ are known.  If either one is infinite, 
some spin-spin correlation does not decay to zero as $r 
\to \infty$, which is equivalent to the existence of {\em long-range 
order}.

\subsection{Specific operators} \label{sec:4stateOperators}
Now we are ready to predict (in terms of $\KE$ and $\KO$) the 
exponents $\etaM$ and $\etaS$ 
for the magnetization and staggered magnetization operators in 
the 4-state CPAFM.  We shall represent the spins A, B, C, and D as 
unit vectors $\magMoment(\vecx)$ pointing toward the corners of a 
regular tetrahedron.  The staggered magnetization 
$\stagMagMoment(\vecx)$ is defined just as in 
Sec.~\ref{sec:3stateOperators}.  Determining the wavevectors 
associated with the operators $\magMoment(\vecx)$ and 
$\stagMagMoment(\vecx)$ is trickier for the 4-state CPAFM than for the 
3-state Potts AFM because of the higher dimensionality of height 
space.  In addition, the optimal wavevector $\vecG$, 
the one giving the minimal value of $\eta_{\vecG}$,
depends on the ratio $\KE/\KO$.

From equations \eqref{Kmatrix} and \eqref{genEtaG} it is clear that 
the optimal $\vecG$ may be purely even (\ie{}, with no component in 
the odd height space), purely odd, or `mixed' (having both even and 
odd components), depending on whether $\KE/\KO$ is large, small, or 
intermediate in value.  The minimal length reciprocal lattice vector 
$\vecG$ in the purely even, purely odd, or mixed case is
\begin{equation}
\GE = \evenv{\frac{\pi}{2},0,0}, \quad \GO = \oddv{0,\frac{\pi}{3}}, 
	\quad \text{or} \quad \Gmix = \evenv{\frac{\pi}{4},0,\frac{\pi}{4}} + 
	\oddv{0,\frac{\pi}{6}},
\end{equation}
respectively.  Each of these is 
symmetry-related to other $\vecG$s with the same length.  As 
it turns out, the staggered magnetization operator 
$\stagMagMoment(\vecx)$ has nonzero Fourier components at $\Gmix$ 
and $\GE$, while the magnetization $\magMoment(\vecx)$ has nonzero 
Fourier components at $\Gmix$ and $\GO$.

Eq.~\eqref{genEtaG} gives the corresponding exponents,
\begin{equation} \label{KsToEtas}
\etaE = \frac{\pi}{8\KE}, \qquad \etaO = \frac{2\pi}{27\KO}, \qquad
	\etamix = \frac{1}{2}\etaE + \frac{1}{4}\etaO.
\end{equation}
(Since $\etaE$ and $\etaO$ are independent of the way we parameterized 
height space, we shall use them as surrogates for $\KE$ and $\KO$ in 
subsequent expressions.)
The exponent in \eqref{genEta} is $\etamin = \min\left\{ \etaE, 
\etaO, \etamix \right\}$.  In particular,
\begin{mysubequations}
\begin{gather}
\text{if } \etarat < 1/2, \text{ then } \etamin = \etaE \text{ and } 
	\etaM > \etaS = \etamin = \etaE,\\
\text{if } 1/2 \leq \etarat \leq 3/2, \text{ then } \etamin = \etamix 
	\text{ and } \etaM = \etaS = \etamin = \etamix,\\
\text{and if } \etarat > 3/2, \text{ then } \etamin = \etaO 
	\text{ and } \etaS > \etaM = \etamin = \etaO.
\end{gather}
\end{mysubequations}%

\subsection{Locking exponent} \label{sec:4stateLock}
Roughly speaking, there are more states with coarse-grained $\vech$ 
near a node 
than near an interstitial 
point of the ideal-states graph. In effect, an 
operator 
$\flock(\vech)$
is added to the effective free energy,  
a periodic function of $\vech$ which is negative 
near each node of the graph. 
This (small) term favors `locking' 
of the entire system into a `smooth' state, which is close to
one of the ideal states. 
It is known that locking occurs whenever the corresponding 
exponent $\etalock$ is less than 4~\cite{JoseEtal:1977}.

This forces our attention to the Bravais lattice of the ideal-states 
graph. Usually this is denser than the repeat lattice, since it 
includes translations which make the nodes superpose, even if they 
represent different ideal states.  Correspondingly, $\Glock$, the 
smallest reciprocal lattice vector of the operator $\flock$, is 
typically larger than $\Gmin$.  Thus, in the 4-state CPAFM case, the 
`even' projection of the ideal-states graph has a bcc Bravais lattice, 
which has twice as many points as the simple cubic repeat lattice, and 
correspondingly has $\GlockE$ larger than $\GE$ defined in
Sec.~\ref{sec:4stateExponents}.  However, when the graph is 
considered in $5$ dimensions, it turns out that its Bravais lattice is 
the same as the repeat lattice, since the `odd' components deviate in 
the opposite directions from the body-corner and body-center points 
of the `even' projection.  Although the corresponding reciprocal 
lattice contains $\GE$, there is an extinction in the Fourier 
transform of the ideal-states graph and thus $\flock(\vech)$ does not 
have a component at wavevector $\GE$.

We find that the optimal reciprocal lattice vector is
\begin{equation}
\GlockE = \evenv{\frac{\pi}{2},\frac{\pi}{2},0}, \quad
\GlockO = \oddv{0,\frac{2\pi}{3}}, \quad \text{or} \quad 
\Glockmix = \evenv{\frac{\pi}{4},\frac{\pi}{4},0} + 
	\oddv{\frac{\pi}{6},\frac{\pi}{6}}.
\end{equation}
(Each of these is symmetry-related to other $\vecG$s with the 
same length).
Consequently, in terms of $\etaE$ and $\etaO$ 
as used in the preceding section,
\begin{mysubequations}
\begin{gather}
\text{if } \etarat < 1/6, \text{ then } \etalock = \etalockE = 
	2\etaE,\\
\text{if } 1/6 \leq \etarat \leq 15/2, \text{ then } \etalock = 
	\etalockmix = \frac{1}{4}\etalockE + 
	\frac{1}{16}\etalockO,\label{etaLockMidRange}\\
\text{and if } \etarat > 15/2, \text{ then } \etalock = \etalockO = 
	4\etaO.
\end{gather}
\end{mysubequations}%

%% file: CPAFM4.tex
\section{Monte Carlo simulation} \label{sec:4}
To obtain more information on the 4-state CPAFM we performed a Monte 
Carlo simulation of the model.  By averaging appropriate quantities 
over the ensemble of allowed states, we can independently determine 
the stiffnesses, $\KE$ and $\KO$, and any correlation exponents.  The 
results can then be examined in light of Section~\ref{sec:3}.

Although we are only particularly interested in the case $q = 4$, we 
developed an algorithm for simulating the general $q$-state CPAFM.  
This permitted us to perform a partial check of our code by simulating 
the $q = 3$ Potts AFM and comparing our Monte Carlo results with the 
known exact results for that model.

\subsection{Monte Carlo algorithm} \label{sec:mCAlgorithm}
We chose to use periodic boundary conditions in the simulation.  Given 
this choice, it is natural to require that the number of rows and 
columns in the lattice both be even numbers; otherwise, there can be 
no well-defined even and odd sublattices and, hence, the odd heights 
are not well-defined.  In addition, there would be no allowed 
configuration corresponding to a single ideal state.

\subsubsection{Initial state}
We performed runs with each of three types of initial state.  The 
simplest choice is one of the $2q$ ideal states.  All the spins in 
one sublattice are set to the same value, while the spins in the 
other, disordered sublattice are chosen randomly from the remaining 
$q-1$ possible values.  The most important property of this state is 
that it is `flat' in height space (the coarse-grained height is 
constant throughout the system).  As the system evolves into a rough 
state, the Fourier spectrum $\langle |\vech(\vecq)|^{2} \rangle$ 
increases from zero to the equilibrium value, for most $\vecq$.

A second, more elaborate method is to divide the lattice into an array 
of smaller blocks; each block is assigned a random ideal state (to be 
filled in as described above), with certain restrictions to ensure 
that only allowed configurations are obtained for the full lattice.  
This initial state is also macroscopically flat.

A third, contrasting kind of initial state for a height model is not 
macroscopically flat; it has, locally, the maximum possible 
height-space slope~\cite{KondevHenley:1995a,KondevHenley:1995b,%
RaghavanEtal:1997}.  It is called a `roof' configuration since it is 
usually composed of two domains of opposite slope.  (The 
system-average slope must be zero in order for the states of lowest 
free energy to be accessible by local updates.)  The `roof' 
configuration shown in Figure~\ref{fig:roof}
\begin{figure}\tt
\settowidth{\figboxwidth}{A B C A B A C B A C}
\addtolength{\figboxwidth}{10mm}
\parbox[t]{\figboxwidth}{(a)\\
\\
A B C A B A C B A C\\
B C A B C B A C B A\\
C A B C A C B A C B\\
A B C A B A C B A C\\
B C A B C B A C B A\\
A B C A B A C B A C\\
C A B C A C B A C B\\
B C A B C B A C B A\\
A B C A B A C B A C\\
C A B C A C B A C B}
\parbox[t]{\figboxwidth}{(b)\\
\\
6 6 5 4 4 4 5 6 6 7\\
6 5 4 4 3 4 4 5 6 6\\
5 4 4 3 2 3 4 4 5 6\\
4 4 3 2 2 2 3 4 4 5\\
4 3 2 2 1 2 2 3 4 4\\
4 4 3 2 2 2 3 4 4 5\\
5 4 4 3 2 3 4 4 5 6\\
6 5 4 4 3 4 4 5 6 6\\
6 6 5 4 4 4 5 6 6 7\\
7 6 6 5 4 5 6 6 7 8}
\parbox[t]{\figboxwidth}{(c)\\
\\
1 3 4 3 1 3 4 3 1 0\\
3 4 3 1 0 1 3 4 3 1\\
4 3 1 0 1 0 1 3 4 3\\
3 1 0 1 3 1 0 1 3 4\\
1 0 1 3 4 3 1 0 1 3\\
3 1 0 1 3 1 0 1 3 4\\
4 3 1 0 1 0 1 3 4 3\\
3 4 3 1 0 1 3 4 3 1\\
1 3 4 3 1 3 4 3 1 0\\
0 1 3 4 3 4 3 1 0 1}
	\sepCaption{(a) A `roof' starting configuration for the 3-state 
	Potts AFM.  The height variables in each quadrant have a (different) 
	uniform tilt, which is the largest possible.  This is also a valid 
	`roof' configuration for the 4-state CPAFM model.
	(b) The corresponding configuration of even height component $h_{1}$ 
	in the 4-state CPAFM.
	(c) The corresponding configuration of odd height component $h_{4}$
	in the 4-state CPAFM.}
	\label{fig:roof}
\end{figure}
includes 4 domains of periodic states.  In each domain (considered as 
a state in the $q = 4$ model), the odd heights are flat, while the 
even heights vary as rapidly as allowed.  In this case, many Fourier 
amplitudes are large at first and approach their equilibrium values 
from above.  Thus, by iterating until the expectation value is 
independent of initial configuration, we can ensure that equilibrium 
is reached.

\subsubsection{Update move} \label{sec:updateMove}
An update move must satisfy three conditions:
\begin{enumerate}
	\item[(i)]  it must change any allowed configuration into another allowed 
	configuration;
	\item[(ii)]  it must be ergodic -- \ie{}, through a sequence of update 
	moves, any allowed configuration must be accessible from any other 
	one; and
	\item[(iii)]  it must satisfy detailed balance.
\end{enumerate}

In the CPAFM model the first condition is nontrivial: a single 
spin-flip is disallowed on many sites because of the nearest-neighbor 
and plaquette constraints, and the remaining allowed single-spin flips 
are too few to be ergodic.  Thus a cluster-update move is a {\em 
necessity} for the CPAFM ground-state ensemble.\footnote{In the 
3-state Potts AFM, it is an optional improvement to obtain 
acceleration ~\cite{WangEtal:1990}.}
Clearly, within a cluster, the change should be a global symmetry of 
the model, \ie{} an exchange of Potts states $X$ and $Y$.  (An allowed 
single-spin flip is a special case of this.)

The update move we used is to
\begin{enumerate}
	\item[(1)] Pick a site $\vecx$ at random; say $\sigma(\vecx) = X$, 
	and randomly choose $Y$ ($\neq{}X$) to be any of the $q-1$ values of 
	spin.
	\item[(2)] Determine the minimal domain containing site $\vecx$ in 
	which it is allowable to exchange $X \leftrightarrow Y$ on all sites 
	(we call this the `$XY$-cluster' of site $\vecx$), and do so.
\end{enumerate}

The geometrical rules for an $XY$-cluster are simple.  We define a 
link (in the percolation sense) between sites $\vecx$ and $\vecx'$ if 
either (i) they are nearest neighbors and have spin values $X$ and 
$Y$; or (ii) they are second-neighbors, both with spin $X$ or both 
with spin $Y$, and the other two spins on the plaquette are 
different.  Then the $XY$-cluster consists of all sites connected 
to $\vecx$ by such links; in this fashion all $X$ and $Y$ spins are 
divided into disjoint clusters, as in a percolation model.  Clearly, 
performing $X \leftrightarrow Y$ on one endpoint of a link and not 
the other creates a disallowed state, so we must at least update the 
$XY$-cluster; it is easy to see that it is also sufficient to do so.

Since the exchange $X \leftrightarrow Y$ does not change the pattern 
of $XY$-clusters, the update move is reversible: it can change 
configuration $\Phi$ directly into configuration $\Psi$ if and only 
if it can change $\Psi$ directly into $\Phi$.  Furthermore, the 
probability of flipping a given $XY$-cluster in an update step is 
just $1/(q-1)$ times the fraction of all lattice sites which belong to 
that cluster (since we would have flipped the same cluster if we had 
initially hit on any of its sites and chosen the appropriate second 
spin value).  Hence the rate of $\Phi \to \Psi$, given 
configuration $\Phi$, is the same as the rate $\Psi \to \Phi$, given 
$\Psi$; for our ensemble, in which all allowed configurations have 
equal weight, this is the detailed balance property.  The only thing 
left is to verify the ergodicity of the update move, which is done in 
Appendix~\ref{sec:appB}.

In speaking of ergodicity, we should make clear that, in the presence 
of periodic boundary conditions, any {\em local} update move 
conserves the net height difference (tilt) across the system and 
hence is non-ergodic in a trivial sense.  However, our cluster update 
move is {\em nonlocal} and can change the net tilt.

It is interesting to note that, in terms of the height variables 
$\vecz(\vecx)$, the update move simply inverts the even height 
components on the sites within the updated cluster (the cluster 
boundary is a loop of constant height, as studied by Kondev and 
Henley~\shortcite{KondevHenley:1995a}, so the height is unchanged 
along it).  This reflection is closely related to the 
`valleys-to-mountains reflection' algorithm~\cite{EvertzEtal:1991,%
HasenbuschEtal:1992}.  On the other hand, the odd height components 
on the cluster sites are simply shifted by a constant vector.

\subsection{Measurements}
Simulation runs were performed for lattices of various sizes ranging 
from $16 \times 16$ to $200 \times 200$.  As mentioned above, since 
the notion of even and odd sublattices is essential in these models, 
the number of sites in a row or column of the lattice was always 
chosen to be even.  In runs where Fourier transforms were taken, fast 
Fourier transform (FFT) algorithms were used and, hence, it was also 
convenient in these runs to have the number of sites in a row or 
column be a power of two.

Since successive states are highly correlated, any measurements were 
taken once per sampling interval; the duration of a sampling interval 
(as a number of cluster hits) was varied depending on the lattice 
size.  Furthermore, in order to estimate uncertainties in the 
calculated averages, we computed averages and variances using groups 
of $100$ sampling intervals.  Averages over successive groups were 
treated as uncorrelated for the purpose of estimating uncertainties.

We made the following measurements.  At each sampling time, the 
height configuration $\vecz(\vecx)$ was generated from the spin 
configuration.  Then $\vech(\vecq)$, the Fourier transform of the 
height configuration $\vecz(\vecx)$, was computed (via FFT).  The 
average of $|\vech(\vecq)|^{2}$ was accumulated for every 
wavevector $\vecq$.

We must note that our procedure in generating $\vecz(\vecx)$ is not 
quite safe.  The FFT requires that $\vecz(\vecx)$ obey periodic 
boundary conditions; but if the configuration has a net tilt in 
height space, the algorithm leaves a `seam' along 
the edges of the lattice, across which $\vecz(\vecx)$ has a sudden 
jump.  (Recall that the tilt can be changed by the algorithm only in 
the rare event than an update $XY$-cluster winds nontrivially around 
the periodic boundaries.)  If `tilted' configurations are frequent, 
this causes a spurious contribution to $\langle |\vech(\vecq)|^{2} 
\rangle$ for (only) $\vecq$ lying on the $x$ and $y$ axes of Fourier 
space.\footnote{Such behavior was very obvious in (unpublished) test runs 
of Kondev and Henley~\shortcite{KondevHenley:1995b}. Their published 
data were from simulations that disallowed any update moves that would 
have changed the tilt.}
That would be visible in data plots such as
Figure~\ref{fig:4stateStraightPlot}: those data 
points would form a separate curve, shifted above the points from 
elsewhere in Fourier space.  In fact, the plotted points appear to 
fall along a single curve.  Therefore we believe that configurations 
with height-space tilt are rare and do not affect our results.  (If 
the system is in a `smooth' phase, as discussed in Sec.~\ref{sec:5}, 
suppression of tilt is just what one would expect.)

For the 4-state model, additional measurements were made.  We computed 
the total magnetization $\magnetization = \sum \magMoment(\vecx)$ and 
staggered magnetization $\stagMag = \sum \stagMagMoment(\vecx)$, and 
accumulated averages of $|\magnetization|^{2}$ and $|\stagMag|^{2}$.  
Also, we measured the size distribution of the $XY$-clusters.

%% file: CPAFM5.tex
\section{Simulation results and discussion} \label{sec:5}
We will now present and discuss the Monte Carlo data.  Our chief 
interest is whether the heights are in a smooth or a rough phase, 
corresponding respectively to long-range order or power-law 
correlations of the underlying Potts spins.  Since the `even' and 
`odd' height components are not related by symmetry, it is also 
possible that one kind could be smooth and the other kind could be 
rough, as occurs in the dimer-loop model~\cite{RaghavanEtal:1997}.

The `smooth' scenario is favored by measurements of the height 
fluctuations, which contain the most information
(Secs.~\ref{sec:fourierFlucts} and \ref{sec:evenOddFlucts}).  On the 
other hand, direct measurements (Sec.~\ref{sec:sizeScaling}) of the 
order parameters suggest a `rough' phase.  The evidence will be 
judged in Sec.~\ref{sec:roughOrSmooth}.

As a test, we simulated the 3-state Potts AFM (using the same code, 
as explained in Sec.~\ref{sec:4}), so as to compare with exact 
results.  (The 3-state model, in its representation as the BCSOS 
model, was also used for test simulations by Raghavan 
\etal{}~\shortcite{RaghavanEtal:1997}.)  According to the theory in 
Sec.~\ref{sec:2}, $\lim_{q \to 0} |\vecq|^{2} \langle |h(\vecq)|^{2} 
\rangle = 1/K$.  In Figure~\ref{fig:3stateStraightPlot}
\begin{figure}
\epsfig{file=fig05.epsi,angle=-90, width=5.5in}
	\sepCaption{Plot of $|\vecq|^{2} \langle |h(\vecq)|^{2} \rangle$ vs. 
	$|\vecq|^{2}$ for the 3-state simulation (size $64 \times 64$).  The 
	(nearly) horizontal line shows the best linear fit to the data.}
	\label{fig:3stateStraightPlot}
\end{figure}
we show $|\vecq|^{2} \langle |h(\vecq)|^{2} \rangle$ vs. 
$|\vecq|^{2}$ for the 3-state simulation on a $64 \times 64$ 
lattice.  Clearly this quantity does approach a constant as $|\vecq| 
\to 0$.  Although the statistical uncertainty in each data point is 
$\approx 10\%$, the uncertainty of the intercept in 
Figure~\ref{fig:3stateStraightPlot} is only $\approx 1\%$ according 
to a least squares fit.  The fitted inverse stiffness constant is 
$1/K = 1.92 \pm 0.02$, which is in excellent agreement with the known 
exact value of $6/\pi \approx 1.91$.

\subsection{Fourier mode fluctuations} \label{sec:fourierFlucts}
The expected behavior of the height components, in the limit $q \to 0$ 
and $L \to \infty$, can be put in the form
\begin{equation} \label{limitingFlucts}
\langle |\htilde_{i}(\vecq)|^{2} \rangle \sim \frac{C_{i}}{q^{2-y_{i}}}.
\end{equation}
When the height components are rough and described by a 
gradient-squared free energy, Eq.~\ref{genFluctuations}, then 
$y_{i} = 0$ and $C_{i} = \left(\tensorK^{-1}\right)_{ii}$.  If we 
define height components to be `smooth' either when
$\left(\tensorK^{-1}\right)_{ii} = 0$, or when the height 
fluctuations are bounded independent of system size, this is 
equivalent to $y_{i} > 0$ in \eqref{limitingFlucts}.  In such a case, 
the boundedness of height fluctuations, $\langle |\vech - \vech'|^{2} 
\rangle < \const$, implies, when inserted in \eqref{oneDAlgDecay}, 
that operator correlations $\langle \opO(\vecx) \opO(\vecx') \rangle$ 
are bounded below as $|\vecx - \vecx'| \to \infty$.  One's usual 
picture of a `smooth' phase is a background of long-range order, on 
which some height fluctuations with rapidly decaying correlations are 
superposed.  That would imply $q$-independent behavior of
\eqref{limitingFlucts} at long wavelengths, \ie{} $y = 2$.

Figure~\ref{fig:4stateStraightPlot} shows $\log\left(q^{2} \langle 
|h_{1}(\vecq)|^{2} \rangle\right)$ vs. $\log\left(q^{2}\right)$ for 
the 4-state simulation on a $64 \times 64$ lattice.
\begin{figure}
\epsfig{file=fig06.epsi,angle=-90, width=5.5in}
	\sepCaption{Plot of $|\vecq|^{2} \langle |h_{1}(\vecq)|^{2} \rangle$ vs. 
	$|\vecq|^{2}$ (log-log) for the 4-state CPAFM model.}
	\label{fig:4stateStraightPlot}
\end{figure}
The slope of the graph at small $q$ is $y_{1}/2$ and its evident 
nonzero value implies that $h_{1}$ is smooth.  A linear fit to the 
small-$q$ data gave
\begin{equation} \label{fluctExp}
y_{1} = 0.27 \pm 0.02,
\end{equation}
where the uncertainty reflects the dependence on choosing an upper 
cut-off $q$ for data points to be used in the fit.  The data for 
$h_{2}$ and $h_{3}$ (not shown) indicate similar values for $y_{2}$ 
and $y_{3}$, as expected from the symmetry among all even height 
components.

As for the `odd' height components, we find numerically for all 
$\vecq$ ($\neq 0$) that
\begin{equation} \label{oddEvenEquality}
\langle |\vechodd(\vecq)|^{2} \rangle = 
						3 \langle |\vecheven(\vecq)|^{2} \rangle.
\end{equation}
This is seen in the nice overlap of plots of
$q^{2} \langle |h_{1}(\vecq)|^{2} \rangle$ and 
$q^{2} \langle |h_{4}(\vecq)|^{2} \rangle/3$ against 
$q^{2}$(Figure~\ref{fig:4stateOddEvenOverlay}).
\begin{figure}
\epsfig{file=fig07.epsi,angle=-90, width=5.5in}
	\sepCaption{This graph overlays plots of 
	$|\vecq|^{2} \langle |h_{1}(\vecq)|^{2} \rangle$ and of 
	$|\vecq|^{2} \langle |h_{4}(\vecq)|^{2} \rangle/3$,  
                against $|\vecq|^{2}$ for the 4-state CPAFM model 
         showing they cannot be distinguished.
     (Data are from a $64 \times 64$ lattice and,  for clarity, 
      only $\vecq \parallel
        (\pm 1, \pm  1)$ have been included.)}
	\label{fig:4stateOddEvenOverlay}
\end{figure}
The corollary of \eqref{oddEvenEquality} is that $y_{i}$ has the same 
value for all height components.  (The significance of \eqref{oddEvenEquality} 
is discussed in Sec.~\ref{sec:evenOddFlucts}.)  Hence all height 
components are `smooth' and we expect the spins to exhibit long-range 
order.

However, as noted above, the expectation for a smooth phase would be 
$y_{i} = 2$, contrary to \eqref{fluctExp}.  Therefore, we carried out 
a finite size analysis to check whether the low-$q$ behavior was a 
system-size effect.  For nonzero $q$ and finite $L$ we expect
\eqref{limitingFlucts} to generalize to
\begin{equation}
\langle |\htilde_{i}(\vecq)|^{2} \rangle = q^{y_{i}-2} f_{i}(q,L),
\end{equation}
where $f_{i}(q,L)$ is a scaling function that approaches $C_{i}$ as 
$q \to 0$ and $L \to \infty$.  Figure~\ref{fig:4stateScalingPlot} shows a plot
\begin{figure}
\epsfig{file=fig08.epsi,angle=-90, width=5.5in}
	\sepCaption{Scaling plot overlaying data on
	$|\vecq|^{2} \langle |h_{1}(\vecq)|^{2} \rangle$ vs. $|\vecq|^{2}$ 
         for the 4-state CPAFM model, for various system sizes.
         (The largest $\vecq|$ values have been omitted.)}
	\label{fig:4stateScalingPlot}
\end{figure}
of $q^{2} \langle |h_{1}(\vecq)|^{2} \rangle$ vs. $q^{2}$ for $16 
\times 16$, $32 \times 32$, and $64 \times 64$ lattices.  The nice 
overlap of the three functions indicates that $L$ in these runs is 
sufficiently large that $f_{i}(q,L)$ is a function of $q$ only.

The finite size of $L$ also imposes a limit on the smallest nonzero 
wavevectors $\vecq$ which can be probed in a simulation.  But the plot 
in Figure~\ref{fig:4stateStraightPlot} appears rather linear (as 
already noted, its slope varies by only $\pm 0.02$ over a large range 
of $q$ values), which would suggest that the $q$ values accessible in 
the $L = 64$ simulation are small enough that $f_{i}(q,L)$ is close 
to $f_{i}(0,L)$, and hence that the value extracted for $y_{i}$ in 
\eqref{fluctExp} is close to the correct asymptotic one.

\subsection{Ratio of `even' and `odd' height fluctuations} 
\label{sec:evenOddFlucts}
The ratio of $3$ in Eq.~\eqref{oddEvenEquality} is nontrivial.  It is 
certainly not true that $|\vechodd(\vecq)|^{2} = 3 |\vecheven(\vecq)|^{2}$ 
for every allowed configuration; indeed, Figure~\ref{fig:roof} shows a 
case where $|\vechodd(\vecq)|^{2}$ is (essentially) zero while 
$|\vecheven(\vecq)|^{2}$ is unusually large for $\vecq = 
\left(0,\pm\pi/L\right)$ or $\left(\pm\pi/L,0\right)$.  Nor could we 
find any hidden symmetry relating the `even' and `odd' height 
components.

There is a simple explanation for the observation 
\eqref{oddEvenEquality} if we postulate that
\begin{enumerate}
	\item[(i)] both even and odd height components are `smooth', and
	\item[(ii)] the system is dominated by domains of a single ideal 
	state, with small independent domains of the three other ideal 
	states which are most similar to it.
\end{enumerate}

In the ideal-states graph, those neighboring ideal states have 
vertices surrounding that of the dominant state, forming $120^{\circ}$ 
angles.  (The same angles may be seen in both the even and odd 
projections of the ideal-states graph.)  Consequently the height 
fluctuations all occur in a rank-$2$ subspace of height space and in 
particular, $\vecheven(\vecq)$ is a linear multiple of 
$\vechodd(\vecq)$.  The linear ratio is in fact $1/\sqrt{3}$ -- the 
ratio of the even and odd components in 
Table~\ref{table:heightSteps} -- and \eqref{oddEvenEquality} follows 
immediately.

In fact, if the fluctuations occurred around a single ideal state, 
there would be cross-correlations between $\vecheven(\vecq)$ and 
$\vechodd(\vecq)$.  However, thanks to the cluster update move, the 
dominant domain is frequently being changed; thus cross-correlations 
are averaged to zero.

In fact, we can even extend this argument to permit domains which 
are {\em two} steps away from `AX' on the ideal-states graph.  The 
second neighbor distances on the ideal-states graph are also in the 
ratio $1/\sqrt{3}$, so that the contributions {\em of any single 
domain} to $\vecheven(\vecq)$ and $\vechodd(\vecq)$ are still in the 
right ratio.  However, the different domains do span a space of rank 
greater than $2$, so the total $\vecheven(\vecq)$ and 
$\vechodd(\vecq)$ are no longer linear multiples.  Nevertheless, if 
the locations of the different domains are statistically independent, 
then cross-terms between different domains must cancel and one again 
obtains \eqref{oddEvenEquality}.  Because small inter-domain 
correlations must exist, and there must be rare domains which are 
three steps away from the dominant one (third neighbor distances on 
the ideal-states graph are not in the right ratio), we conjecture that, 
in fact, there are tiny deviations from an exact ratio of $3$ in 
\eqref{oddEvenEquality}.

Thus, the observed ratio of $3$ implies bounded fluctuations, which 
imply long-range order.  Indeed, the only way we expect this behavior 
is when a finite fraction of the allowed configurations contain an AX 
domain that occupies a finite fraction of the lattice.

\subsection{System-size scaling of operators} \label{sec:sizeScaling}
Say an operator $\opO(\vecx)$ has correlations 
$\langle \opO(\vecx) \opO(\vecx') \rangle \sim r^{-\eta}$, and the 
corresponding order parameter is defined as
$\displaystyle M_{\opO} = \sum_{\vecx} 
\opO(\vecx)$.  Then simply expanding the double sum in $\langle M_{\opO} 
M_{\opO} \rangle$ for a system of $N = L^{2}$ sites gives
\begin{equation}
\label{eq:MOexpect}
\langle |M_{\opO}|^{2} \rangle \sim L^{4-\eta} \sim N^{2-\eta/2}.
\end{equation}
In particular, $\langle |M_{\opO}|^{2} \rangle \sim N^{2}$ in a system 
with long-range order.
We measured expectation \eqref{eq:MOexpect} for 
$M_{\opO} \to \magnetization$ and 
$M_{\opO} \to \stagMag$, corresponding to 
exponents $\etaM$ and $\etaS$ defined in Sec.~\ref{sec:4stateOperators}.

Figure~\ref{fig:4stateMagPlot} shows a log-log plot of 
\begin{figure}
\epsfig{file=fig09.epsi,angle=-90, width=5.5in}
	\sepCaption{Mean square total magnetization 
	$\langle |\magnetization|^{2} \rangle$ vs. system size $N$ (log-log 
	plot).  The line is the fit $\langle |\magnetization|^{2} \rangle 
	\sim N^{1.85}$.}
	\label{fig:4stateMagPlot}
\end{figure}
$\langle |\magnetization|^{2} \rangle$ 
from the 4-state simulations with three system sizes $N$. 
The data points
fit to a line with slope $1.85 \pm 0.01$ (fitting error only!),  
\ie 
\begin{equation} \label{magFit}
\langle |\magnetization|^{2} \rangle \sim N^{1.85}
\end{equation}
for $N$ in the range simulated.

We also measured the average total staggered magnetization for 
lattices of size $64 \times 64$ and $120 \times 120$; connecting these 
two data points would imply
\begin{equation} \label{stagMagFit}
\langle |\stagMag|^{2} \rangle \sim N^{1.93}.
\end{equation}
Eqs.~\eqref{magFit} and \eqref{stagMagFit} respectively imply
\begin{equation} \label{experEtas}
\etaM \approx 0.30 \pm 0.02,  \qquad \etaS \approx 0.14.
\end{equation}

\subsection{Discussion: rough or smooth?} \label{sec:roughOrSmooth}
The magnetization behavior reported in Sec.~\ref{sec:sizeScaling} is 
that expected for rough heights and critical spin correlations, which 
is in contradiction to our conclusion from Fourier fluctuations in 
Sec.~\ref{sec:fourierFlucts}.  Which inference should we believe?  We 
shall attempt to make sense of the data assuming a `rough' state, and 
show it leads to several contradictions.

We start with the $\vech(\vecq)$ data.  The ratio of $3$ in 
\eqref{oddEvenEquality}, which is our surest and most precise 
numerical result, implies that $\KO = \frac{4}{9}\KE$.  Then, in the 
notation of Sec.~\ref{sec:4stateOperators}, it would follow that
\begin{equation}
\etaE = \frac{\pi}{8\KE}, \qquad \etaO = \frac{\pi}{6\KE},
	\qquad \etamix = \frac{5\pi}{48\KE},
\end{equation}
and hence
\begin{equation} \label{oddEvenRatioEtas}
\etaM = \etaS = \etamix.
\end{equation}
Roughness means that the curve in Figure~\ref{fig:4stateStraightPlot} 
asymptotes to a constant as $q \to 0$; we may assume it does so 
monotonically, so the smallest visible value provides a lower bound on 
the stiffnesses, $\KE \ge 1.5$, \ie{}
\begin{equation} \label{etaBound}
\etamix \le 0.22.
\end{equation}

The conclusion $\etaM = \etaS$ in Eq.~\eqref{oddEvenRatioEtas} is 
incompatible with the unequal values obtained in 
Eq.~\eqref{experEtas}; furthermore, the value $\etaM \approx 0.30$ 
found from the magnetization data violates the inequality \eqref{etaBound}.

Most convincingly, since $\etarat = 3/4$ we find (from \eqref{KsToEtas} 
and \eqref{etaLockMidRange}) that the locking exponent of 
Sec.~\ref{sec:4stateLock} is $\etalock = \etalockmix \le 0.2$, far 
smaller than the critical value $4$.  Then~\cite{JoseEtal:1977} the 
effective stiffness constant becomes larger and larger on longer 
length scales, so the system must lock into a `smooth' phase.  So, 
based on all the above arguments, we conclude that the system must be 
in a `smooth' phase.

We can check this picture by examining the system in real space.  
Long-range order, in which the system locks into a particular ideal 
state, means that a typical configuration should look like an ideal 
state with some disorder: that is, there should be an excess of one 
particular spin label on one sublattice and a paucity of that spin 
label on the other sublattice, even in an infinite lattice.

A measurement of this tendency can be gleaned from our results on the 
size distribution of $XY$-clusters (these were defined in 
Sec.~\ref{sec:updateMove}).  A substantial fraction of the 
encountered configurations had an $XY$-cluster which included more 
than half of all sites; the frequency of such configurations 
decreases with lattice size, but is still quite appreciable at our 
largest lattice size, $200 \times 200$.  Thus we cannot decide 
whether or not the frequency is zero in the thermodynamic limit.

To further illustrate the tendency to long-range order, 
Figure~\ref{fig:typConfig} shows a typical configuration (after
equilibration) in the 4-state simulation.
\begin{figure}\tt
\parbox[t]{7in}{%
A D A C A B A B C B A D C D B D B D A C A D A B A C A C A B A D\\
D A C A C A B D B A B A D B A B D A C A D C D A B A B A B C B A\\
A C A B A B A B C B A C A D B A B D A D C B C D A C A D A B A C\\
D A C A D A B A B A C A C A D B C B D A D C B C D A C A C A C A\\
C D A D A B A C A C A C B C A D B D A D C D C D A D A C A B A B\\
D A B A B A D A D A B A C A D B D B D C D A D A D A C D C A B A\\
A D A B A D A D A C A D A D A D A D B D A D A D B D A C A D A D\\
D A C A B A B A D A D C D A D C D B D B D B D B D C D A D A D A\\
A B A B A D A C A D B D C D C D B A B D B D B D C D A C A B A B\\
D A D A B A B A B A D B D A D B D B A B D A D B D A C A C A C A\\
A C A D A B A D A D A D C D C D B D B D A B A D B D A C A C A C\\
C A C A B A D A B A B A D A D A D A D B D A C A D A C A B A B A\\
A D A C A C A D A B C B A C A B A B A D A B A B A C B C A C A D\\
C A C A B A D A D A B A B A D A B A D A C A B A D A C A B A B A\\
A D A D A C A D A B A C A C A D A D A D A D A D A B A D A D A B\\
D A B A B A D A B A B A B A C A C A B A D A D A B C B A B A B C\\
A C A C A D A B A B A D A C A B A D A B A B A B C B D B A B A B\\
C A C A B A B A C A D B D A C A D A B A B A B A B A B C B A B A\\
A B A B A C A C A D A D A B A B A D A B D B A B A D A B D B C B\\
B A B A C A D A B A B A D A B A C A D A B A C A B A B A B C B A\\
D B A B A D A C A C A C A B A C A B A D A C A B C B C B D B A D\\
B D B A D A C A B A B A B A D A C A D A C A C A B D B C B C B A\\
D B C B A B A B D B A B A B A B A B A C A C A D A B A B D B C B\\
C D B A B A B D B D B A B D B A B A C A B A D A B C B D B D B D\\
D B C B A D A B C B C B C B A D A C A C A D A C A B A B C B D B\\
B C B A C A C A B D B A B A D B D A D A C A B A D A B C B A B C\\
C B C B A D A D A B A B A D B D C D B D A B A B A B D B C B D B\\
B D B A C A D A C A C A D A D A D C D C D A D A D A B D B C B C\\
D B D B A D A C A B A B A C A D B D A D C D A B A D A B C B D B\\
B D B C B A C A C A D A B A D A D C D A D A D A B A C A B D B A\\
A B A B A B A D A C A D A D C D C D A D A C A B A B A B D B A B\\
B A B C B D B A D A D B D A D A D A C A B A D A D A D A B D B A}
	\sepCaption{A $32 \times 32$ portion of a typical configuration 
	found in simulation of the $L = 64$ system.}
	\label{fig:typConfig}
\end{figure}
Note that roughly $2/3$ of the sites in one sublattice have spin A 
whereas less than $3\%$ of the sites in the other sublattice have 
spin A.  In addition, all of the spin A sites in the ordered 
sublattice have the same height.  The remaining sites are divided 
roughly equally among the remaining spin labels B, C, and D.  These 
are the correlations expected in a long-range-ordered `AX' type 
state.  The fluctuations from the `AX' state (where the first 
sublattice has a spin other than A) appear to be domains of other 
ideal states that are one step from `AX' on the ideal states graph; 
that is consistent with our explanation of the ratio of $3$ in 
Sec~\ref{sec:evenOddFlucts}.

A final check of the long-range order hypothesis is to compute the 
total variance of each height component around its mean, which is 
$\langle |z_{i}(\vecx)|^{2} \rangle = \frac{1}{N} \sum_{\vecq \neq 
0} \langle |h_{i}(\vecq)|^{2} \rangle$.  This comes out to be about 
$0.6$ for the `even' heights, which is also consistent with the 
picture just presented.

As a tentative resolution of the contradictory features of our data, 
we suggest that our model may be close to, or even at, the 
unlocking ($=$roughening) transition.  In that case the correlation 
length would be very large and it would be hopeless to detect the 
asymptotic behavior of the staggered magnetization in accessible 
system sizes.  (We believe that the data in Figure~\ref{fig:4stateMagPlot} 
are not indicative of the asymptotic behavior as $N \to \infty$.)

The anomalous power $y = 0.27$ extracted from the height fluctuation data 
might be explained if the proper functional form were
\begin{equation}
\langle |\vech(\vecq)|^{2} \rangle \sim \log|q_{\nought}/q|/|\vecq|^{2}.
\end{equation}
This functional form is close to that for the `rough' case, but 
strictly speaking the behavior would be `smooth'.  If we define an 
effective exponent $\yeff = 2 - \d{\log\left(\langle |\vech(\vecq)|^{2}
\rangle\right)}/\d{\log q}$, then taking $q_{\nought}$ to be the width 
of the Brillouin zone, we obtain $\yeff = 
1/\log\left(q/q_{\nought}\right) \approx 1/\log(64) = 0.24$, which is 
comparable to the measured value.

%% file: CPAFM6.tex
\section{Conclusions} \label{sec:6}
We have found a spin model which has a height representation but still 
has long-range order in the ground state.  It proved infeasible to 
detect the long-range order through studying fluctuations in the 
magnetization because the necessary system size is too large.  
However, by observing instead the fluctuations in the Fourier 
transformed height, we find conclusive evidence for long-range order 
through simulating systems of reasonable size.

To discuss the possible finite-temperature behavior of this model, 
and its relation to other models, let us display a Hamiltonian which 
has the CPAFM as its (degenerate) ground states:
\begin{multline} \label{hamiltonian}
H = -\frac{1}{2} \sum_{\text{n.n. }\vecx,\vecx'} J_{1} 
		\delta_{\sigma(\vecx),\sigma(\vecx')}
	- \frac{1}{2} \sum_{\text{n.n.n. }\vecx,\vecx''} J_{2}
		\delta_{\sigma(\vecx),\sigma(\vecx'')}\\
	- \sum_{\text{plaq. }\vecx,\vecx',\vecx'',\vecx'''} J_{4}
		\delta_{\sigma(\vecx),\sigma(\vecx'')}
			\delta_{\sigma(\vecx'),\sigma(\vecx''')},
\end{multline}
where the second sum is taken over second nearest neighbors, and the 
last sum is over plaquettes with $\vecx,\vecx',\vecx'',\vecx'''$ in 
cyclic order around the plaquette.  We must take the case $J_{4} = 
-J_{2}$, making the plaquettes of type (ABAB) and (ABCB) degenerate, 
to make the CPAFM states all degenerate and justify their equal 
weights.

At $T>0$, typical configurations violate the CPAFM constraint at rare 
places; this creates dislocation defects in the height field; when 
$\vech(\vecx)$ is followed along a closed loop around that defect it 
is shifted by a height-space Burgers vector $\vecb$ (which is in the repeat 
lattice).  The correlations of these defects are described by 
vortex-vortex (\ie{} `magnetic') type exponents in the Coulomb-gas 
picture~\cite{Nelson:1983,Nienhuis:1987}.  It can be shown that the 
specific heat and inverse correlation length behave as 
$\e^{-cE_{\nought}/T}$, where $E_{\nought}$ is the energy cost of a 
defect, and the coefficient $c$ is related to its Burgers vector and 
the stiffness constant(s)~\cite{HuseRutenberg:1992,Read:1992,%
KondevHenley:1995a,KondevHenley:1995b,KondevHenley:1996}.\footnote{This 
confirms that when Wang \etal~\shortcite{WangEtal:1989,WangEtal:1990} 
fitted the specific heat of the 3-state Potts AFM to the form 
$\e^{C/T^{x}}$, they should have obtained $x=1$~\cite{FerreiraSokal:1995}.}

The best-known height models with critical ground states, in 
particular the 3-state Potts AFM, have stiffness constants $K$ such 
that the defects are unbound, \ie{}, the system is in the same phase 
as the two-dimensional $XY$-ferromagnet above its Kosterlitz-Thouless 
transition temperature.  In such a case, $T = 0$ is a critical point 
with the spin model being disordered at any finite temperature, since 
that means that the fugacity of the height defects, 
$\e^{-E_{\nought}/T}$, is nonzero.  A richer behavior is shown by the 
large-$S$ triangular Ising antiferromagnet: the height field ensemble 
for the ground state is `smooth' (\ie{} the ground state has 
long-range order), and then as $T$ is increased there are two 
transitions at finite $T$, first to a critical state and then to a 
disordered one~\cite{LipowskiEtal:1995}.  Since we find that the CPAFM 
model is `smooth' at $T = 0$, it seems quite plausible that it 
undergoes analogous transitions at $T > 0$.

The Hamiltonian \eqref{hamiltonian} is equivalent to a Boltzmann 
weight (for each possible configuration of Potts spins)
\begin{multline} \label{bWeight}
W = \prod_{\text{n.n. }\vecx,\vecx'} \left(1 -
		x\delta_{\sigma(\vecx),\sigma(\vecx')}\right) \\
	\times
	\prod_{\text{plaq. }\vecx,\vecx',\vecx'',\vecx'''} \left[
		u + v\left(\delta_{\sigma(\vecx),\sigma(\vecx'')} +
					\delta_{\sigma(\vecx'),\sigma(\vecx''')}\right) +
		y\delta_{\sigma(\vecx),\sigma(\vecx'')}
					\delta_{\sigma(\vecx'),\sigma(\vecx''')}
	\right].
\end{multline}
The CPAFM constraints correspond to $x \equiv 1$ and $u \equiv 0$ in 
\eqref{bWeight}; the equal-weighted ensemble means also taking $v = 
-y > 0$.  Nienhuis~\shortcite{Nienhuis:1990} studied the case where 
$y \equiv 0$ (see his Eq. (3)), which is a different weighting of the 
same constrained states.  By mapping it to a loop model, he 
investigated this model with the number of Potts states $q$ as a 
continuous variable.  It was found (see his Eq. (12)) that the model 
is critical when $q \le \qcrit = 3$ (but would have long-range order 
when $q > \qcrit$).  Our Boltzmann weighting, with $y < 0$, disfavors 
the very flat `ABAB' type plaquettes (relative to Nienhuis's 
weighting) and thus should be rougher, with $\qcrit \ge 3$ in our 
case; this is all consistent with our conclusion that the model is 
smooth, but possibly is near the roughening transition.

In Sec.~\ref{sec:5}, we speculated that the CPAFM model might be near 
the unlocking (roughening) transition.  The 3-state Potts AFM on the 
Kagom\'{e} lattice~\cite{HuseRutenberg:1992} and the edge 4-coloring 
of the square lattice~\cite{KondevHenley:1995a,KondevHenley:1995b,%
KondevHenley:1996} are known, in the ensemble where all 
configurations are weighted equally, to be exactly at their roughening 
transitions, \ie{} the locking operator is marginal.  In the latter 
case $\langle |\vech(\vecq)|^{2} \rangle$, as measured in Monte Carlo 
simulations~\cite{KondevHenley:1995b}, shows a clean $1/|\vecq|^{2}$ 
behavior just like a non-marginal rough case.  In the dimer-loop 
model~\cite{RaghavanEtal:1997}, one of the two height components 
shows clean $1/|\vecq|^{2}$ behavior indicating roughness, but the 
other component shows an intermediate power somewhat like our 
findings for the CPAFM model.

\section*{Acknowledgements:}
We would like to thank J.~Kondev, B.~Nienhuis, R.~Raghavan, 
and C.~Zeng for discussions and comments.  This work was supported by 
NSF grant No. DMR-9214943.

%% file: CPAFMappA.tex
\section{Dimensionality of height space} \label{sec:app0}
In this Appendix we show that the height space has dimension 5 as 
asserted in Sec.~\ref{sec:4stateHeightMap}. 
To do this, we first construct the most general mapping of a 
configuration to a height-like vector, such that the height steps
between two sites depends only on their respective spin values 
$\sigma$, $\sigma'$ and on the parity $\parity$ of (say) the first site. 
\footnote{This assumption is justified since the height rule ought
always to have the same lattice periodicity as an ideal state, which 
is a two-sublattice pattern for the CPAFM case.} 
This turns out to give 12 candidate height components. 
Then we show that only five of these are valid, nontrivial height components, 
and finally we show (by examples) that no further reduction of the 
dimensionality is possible. 

Note first that $\Delta\vecz(\sigma,\sigma',\parity)\equiv 
-\Delta\vecz(\sigma',\sigma,-\parity)$
since  the left- and right-hand sides are the height difference of
the steps from $\vecx$ to $\vecx'$ and back again; thus, it suffices to
define $\Delta\vecz$ for $\parity=+1$.
There are $q(q-1)= 12$ possible combinations of $(\sigma,\sigma')$ since 
neighbors cannot have the same spin value; thus all possible functions
form a 12-dimensional linear space, 
labeled by $(X, Y) \in  \{A,B,C,D\}^2$ with $X\neq Y$.

A candidate height component $\zeta_{XY}(\vecx)$
is defined so
$\Delta \zeta_{XY}= +1$
if $(\sigma, \sigma',\parity)=(X,Y, +1)$, 
$\Delta \zeta_{XY}= -1$
if $(\sigma, \sigma',\parity)=(Y,X, -1)$, 
and zero otherwise.
In the CPAFM model, it can be checked that 
every component $\zeta_{XY}$ is well-defined,  by encircling each 
plaquette. (If we encounter an $XY$ step,  we must also encounter 
a $YX$ step,  e.g. in the plaquette $XYXZ$.)

\subsection{Reduction to nontrivial height components}

Some linear combinations of the components 
$\zeta_{XY}(\vecx)$ are bounded so the
corresponding subspaces of the 12-dimensional candidate height space
are trivial. 
Namely, let $\rho_{\sigma,E}(\vecx)=+1$ when $\vecx$ is an even site
and its spin takes the value $\sigma$, and let 
$\rho_{\sigma,E}(\vecx)=0$ otherwise; let 
$\rho_{\sigma,O}(\vecx)$
be the analogous function for odd sites. 
Then $\rho_{\sigma,E}(\vecx) = 
-\sum _{\sigma'\neq \sigma} \zeta_{\sigma\sigma'}(\vecx)$ (modulo a global 
additive constant), 
as can be checked by inspection of the possible $\Delta \rho_{\sigma.E}$ 
values; similarly $\rho_{\sigma,O}(\vecx) =
+\sum _{\sigma'\neq \sigma} \zeta_{\sigma'\sigma}(\vecx)$ (modulo a constant).
There is one linear dependence among the $2q=8$ distinct $\rho$ functions, 
since
$\sum _\sigma (\rho_{\sigma,E}(\vecx) +\rho_{\sigma,O}(\vecx)) = 1.$
Thus there is a seven-dimensional trivial subspace of $\zeta$ space; 
the remaining five dimensions, it can be checked, are spanned by the
five height components defined in Sec.~\ref{sec:4stateHeightMap}. 
(For the $q$-state CPAFM model, this construction would give 
$q^2-3q+1$ dimensions.)

\subsection{Necessity of all five components}

It remains to be shown that there is no further reduction of the height
space dimensionality -- that all five components are necessary. 
The easiest way to exclude the existence of a zero (or bounded and hence
trivial) linear combination is by counterexample: we just 
exhibit five configurations, each of
which has a net gradient only of the corresponding height component. 
We need only consider a one dimensional gradient. 

The bottom half of 
Table~\ref{table:shift4}
shows the height steps corresponding to a four spin sequence
like $(ABCD)$  (assumed followed by another $A$). 
Consider now the 10-spin sequence $(ACBD)(ABC)(ACBD)(ACB)$. 
If the first $(ACBD)$ subsequence starts on an even 
site, then the second $(ACBD)$ starts on an odd site;
hence the even components of the steps are both $\evenv{4,0,0}$ but the
odd components cancel.  Also, the height step of $(ABC)$ (beginning on an 
even site) is canceled (in every component) by the height step of 
$(ACB)$ (beginning on an odd site). 
Thus the net height step is $\evenv{8,0,0}$. A configuration, consisting
of repeats of the 10-spin sequence, has a gradient of only the first even 
height component. 
By use of $(ABCD)$ and $(ACDB)$, 
we can construct a string of spins with 
a gradient of only the second or third even components. 

Furthermore, we could reverse the second copy of the 4-spin
subsequence, obtaining (say) $(ACBD)(ABC)(ADBC)(ACB)$. 
This reverses the height step of the second 4-spin subsequence, so that
now the even components cancel and the odd components add up. 
In this fashion we can construct a string which has a gradient of only
the first or second {\it odd} height component. 
Thus, the five height components used in this paper are nontrivial. 

It should be noted that the above statement concerns only the 
allowed microstates of the system. It is possible -- and indeed, 
likely in models with many height components -- that some of the 
components are ``smooth''. That is, although they {\it can}
have arbitrarily large fluctuations (in an arbitrarily large system), 
the actual fluctuations are essentially bounded due to collective effects. 
Thus, it may be that only a subset of the height components is
described by a nontrivial elastic theory.

%% file: CPAFMappB.tex
\section{Proof for repeat lattice} \label{sec:appA}
In this Appendix we rigorously show that two domains of the same ideal 
state always differ in height by a vector of the repeat lattice, from 
which it follows that the coarse-grained height of any ideal state is 
unique modulo the repeat lattice.  This is done for both cases, the 
3-state Potts AFM (see Sec.~\ref{sec:3stateIdealStates}) and the 
4-state CPAFM model (see Sec.~\ref{sec:4stateRepeatLattice}).

\subsection{Palindromic sequences}
In either case ($q = 3$ or $q = 4$), we focus on the sequence of 
spins $\sigma_{0}\sigma_{1}\sigma_{2}\ldots\sigma_{n-1}\sigma_{n}$ 
connecting two sites.  We do not care which path these spins lie 
along, since the height difference between the first and last site 
depends only on the sequence as in Eq.~\eqref{oneDDeltaH}; or, for 
$q=4$, on the sequence and on the parity $\paro$ of the initial 
site as in Eq.~\eqref{CPAFMDeltaH}.  We denote the height difference 
as 
$\Delta\vecz(\paro;\sigma_{0}\sigma_{1}\sigma_{2}\ldots\sigma_{n})$.  
Now
\begin{multline} \label{spinSequenceSplit}
\Delta\vecz(\paro;\sigma_{0}\sigma_{1}\sigma_{2}\ldots
	\sigma_{j-1}\sigma_{j}\sigma_{j+1}\ldots\sigma_{n-1}\sigma_{n})\\
=
\Delta\vecz(\paro;\sigma_{0}\sigma_{1}\sigma_{2}\ldots
	\sigma_{j-1}\sigma_{j}) +
\Delta\vecz(\parity_{j};\sigma_{j}\sigma_{j+1}\ldots\sigma_{n-1}\sigma_{n}),
\end{multline}
so we shall proceed by breaking all sequences down into a few kinds of 
subsequence.

We call `palindromic' a sequence of form $\sig{0}\sig{1}\ldots\sig{j-1}
\sig{j}\sig{j-1}\ldots\sig{1}\sig{0}$ (with $j \ge 1$), \ie{} the 
sequence has reflection symmetry about one spin.  Of course, the 
corresponding spins both have the same parity.  The net height 
change $\Delta\vecz$ of any palindromic sequence is clearly zero, 
since (as can be checked from the height rule definitions)
$\Delta\vecz(\parity_{i};\sig{i}\sig{i+1}) =
-\Delta\vecz(\parity_{i+1};\sig{i+1}\sig{i})$.
(In fact, this 
obviously must be true since the palindromic sequence also describes a 
path from the initial spin back to itself.)
Therefore, using \eqref{spinSequenceSplit} the net $\Delta\vecz$ of a 
sequence containing this palindrome is invariant if the palindrome is 
reduced to the single spin $\sig{0}$.  Note that such a reduction 
keeps invariant the parity of the length of the sequence.

\subsection{Three-state Potts AFM case}
We are concerned with paths between sites in the same kind of 
domain -- say, an AX domain.  So we limit ourselves to paths which 
connect sites of the same parity.  Then the corresponding sequence of 
spins has an even number of steps, and it begins and ends with the 
same Potts state A.

We can break each such sequence up into subsequences which begin and 
end in A; by \eqref{spinSequenceSplit} the net $\Delta{}z$ is the sum 
of those of the subsequences; furthermore, we can reduce any 
palindromes within the subsequences without affecting this sum.  Thus 
each term in the sum represents a palindrome-free sequence beginning 
and ending with A and containing no interior A's;  the only such 
sequences are A, ABCA, and ACBA, which have $\Delta{}z$ equal to $0$, 
$3$, and $-3$, respectively.  There must be an even number of 
three-step terms, since we chose the total sequence to have an even 
number of steps.  Thus, finally, the net $\Delta{}z$ is a sum of an 
even number of $\pm 3$ terms, and hence is a multiple of $6$, which 
completes the demonstration: the possible height differences 
$\Delta{}h$ between two A~B/C domains are precisely the integer 
multiples of $6$.

Analogous results hold for the macroscopic height difference between 
any two domains of (possibly) different ideal states.  For example, 
Eq.~\eqref{oneDDeltaH} for $\Delta{}h$ between an A~B/C domain and a 
C/A~B domain is simply that for some spin sequence of odd length that 
begins with A and ends with B.  We can always decompose such a 
sequence into two subsequences, the first beginning and ending with A 
and the second beginning with A, ending with B and containing no 
interior A's.  The first sequence has $\Delta{}z \equiv 0$ (mod 6)
and the second subsequence can be reduced to the palindrome-free AB, 
for which $\Delta{}z = 1$.  Hence the macroscopic height difference 
between an A~B/C domain and a C/A~B domain is always congruent to $1$ 
(mod 6).

\subsection{Four-state CPAFM case}
The method of proof for the 3-state model must be extended in this 
case since there are now infinitely many palindrome-free sequences 
beginning and ending with A and containing no interior A's.  To 
address this, let us first replace every spin $\sig{k}$ which is not 
an A spin by the palindrome $\sig{k}A\sig{k}$; clearly this doesn't 
change $\Delta\vecz$.  In this fashion the entire sequence is broken 
into short subsequences and we can write $\Delta\vecz$ as the sum of 
terms of the form $\Delta\vecz(\parity,AXA)$ or 
$\Delta\vecz(\parity,AXYA)$, where $X$ and $Y$ are spins distinct 
from each other and from A.  The palindromes AXA are produced whenever 
the original sequence had an A; such palindromes can be reduced 
leaving only the 3-step terms.  Moreover, since they involve an odd 
number of steps, the parity must alternate between $\parity=1$ and 
$\parity=-1$ in successive terms.  Thus, $\Delta\vecz$ is the sum of 
terms of form $\Delta\vecz(1,AXYA) + \Delta\vecz(-1,AX'Y'A)$, where 
$X$ and $Y$ need not differ from $X'$ and $Y'$.  Constructing all 
possible terms of that form (using Table~\ref{table:shiftvectors}), one 
finds that $\Delta\vecz$ must be a linear combination of the six 
vectors $\evenv{4,0,0}\pm\oddv{0,6}$, $\evenv{0,4,0}\pm\oddv{-6,0}$, 
and $\evenv{0,0,4}\pm\oddv{6,-6}$, which are the same as those in 
Eq.~\eqref{repeatGen}.  (Conversely, it is also obvious that for any 
such linear combination, a sequence exists.)  Clearly the Bravais 
lattice generated by linear combinations of these vectors is the 
`repeat lattice' for the 4-state CPAFM.

%% file: CPAFMappC.tex
\section{Ergodicity of the update move} \label{sec:appB}
Ergodicity requires that it be possible to apply successive update 
moves to any allowed configuration to obtain any other allowed 
configuration.  Since the update move we used is reversible, it is sufficient 
to show that any starting configuration can be converted into the 
configuration with all A's on the even sublattice and all B's on the 
odd sublattice.

Also, note that in an $XY$-cluster all the X's are on one sublattice 
and all the Y's are on the other sublattice.  A corollary is that if a 
site with spin X and a site with spin Y are in the same sublattice, 
then the two sites are not in the same $XY$-cluster.  Similarly, if 
two sites both have spin X but are in different sublattices, then they 
are in different $XY$-clusters for all $Y \neq X$.

Now, consider an arbitrary initial configuration of spins.  Pick any 
site and note its spin, X, and which sublattice it is in.  If the 
chosen site is in the even sublattice and $X \neq$ A, then 
interchange spins $X \leftrightarrow$ A within the $X$A-cluster 
containing the site; if the chosen site is in the odd sublattice and 
$X \neq$ B, then interchange spins $X \leftrightarrow$ B within the 
$X$B-cluster containing the site; otherwise, do not perform an update 
move.  In this way we set the spin on the chosen site to the value it 
takes in the `target' configuration, which consists of A's on the 
even sublattice and B's on the odd sublattice.

In the same manner, proceed one by one through the rest of the sites 
in the lattice, performing the appropriate update move, if necessary, 
to set the site's spin to the value it takes in the target 
configuration.  The corollaries mentioned above ensure that once a 
site's spin has been set to its target value, it is not altered by 
the subsequent update moves used to set other spins to their target 
values; for, a site whose spin is set to its target value and a site 
whose spin is not set to its target value can never be in the same 
update cluster.  For this reason, at the end of the process the target 
configuration has been reached.  Therefore, the update move is ergodic.

Note that this proof is valid whether or not periodic boundary 
conditions are used.  In the case of periodic boundary conditions, 
however, it has been implicitly assumed that the target configuration 
exists, which requires that the lattice have even numbers of rows and 
columns.  As mentioned at the beginning of 
Sec.~\ref{sec:mCAlgorithm}, this requirement is a natural one to 
include in any simulation since it is necessary for the ideal states 
of the model to exist.

Also, note that this proof is equally valid for a standard 
(nonconstrained) Potts AFM so long as `cluster' is appropriately 
redefined.  In fact, the arguments here prove the ergodicity of the 
update move for any bipartite lattice since the target `AB' 
configuration exists.  This does not contradict the statement of 
Lubin and Sokal~\shortcite{LubinSokal:1993} that the algorithm is not 
ergodic on $3m \times 3n$ lattices when $m$ and $n$ are relatively 
prime because we require that $m$ and $n$ both be even.